\documentclass[useAMS,usenatbib]{mn2e}
\usepackage{url}
\usepackage{stmaryrd}
\usepackage[pdftex]{graphicx}
\usepackage[draft]{hyperref}
\usepackage[flushleft]{threeparttable}

\def\aj{AJ}%
\def\apj{ApJ}%
\def\apjs{ApJS}%
\def\aap{A\&A}%
\def\mnras{MNRAS}%
\def\pasp{PASP}%
\def\nat{Nature}%
\def\nELEL{78}
\def\nPP{27}
\def\nELP{71}
\def\nPEL{104}

\def\nsamp{280}
\begin{document}

\title[GAMA Blended Spectra Catalog]{ Galaxy And Mass Assembly (GAMA) Blended Spectra Catalog: 
Strong Galaxy-Galaxy Lens and Occulting Galaxy Pair Candidates.}

\author[B.W. Holwerda]{B.W. Holwerda$^{1}$\thanks{E-mail:
holwerda@strw.leidenuniv.nl,  Twitter: @benneholwerda}, 
I. K. Baldry$^{2}$, 				
M. Alpaslan$^{3}$ 				
A. Bauer$^{4}$, 				
J. Bland-Hawthorn$^{5}$, 		
\newauthor
S. Brough$^{4}$, 				
M. J. I. Brown$^{6}$, 			
M. E. Cluver$^{7}$, 				
C. Conselice$^{8}$				
S.P. Driver$^{9,10}$,			
\newauthor
A.M. Hopkins$^{4}$,  			
D. H. Jones$^{11}$, 				
\'A.R. L\'{o}pez-S\'{a}nchez$^{4,11}$, 
J. Loveday$^{12}$, 				
M.J. Meyer$^{9}$,				
\newauthor
and
A. Moffett$^{9}$\\	
$^{1}$ University of Leiden, Sterrenwacht Leiden, Niels Bohrweg 2, NL-2333 CA Leiden, The Netherlands\\
$^{2}$ Astrophysics Research Institute, Liverpool John Moores University, IC2, Liverpool Science Park, 146 Brownlow Hill, Liverpool, L3 5RF, UK\\
$^{3}$ NASA Ames Research Centre, N232, Moffett Field, Mountain View, CA 94034, USA \\
$^{4}$ Australian Astronomical Observatory, 105 Delhi Rd. North Ryde NSW 2113\\
$^{5}$ Sydney Institute for Astronomy, School of Physics A28, University of Sydney, NSW 2006, Australia\\
$^{6}$ School of Physics, Monash University, Clayton, Vic 3800, Australia \\
$^{7}$ Department of Physics, University of the Western Cape, Robert Sobukwe Road, Bellville, 7530, South Africa \\
$^{8}$ University of Nottingham, School of Physics \& Astronomy, Nottingham, NG7 2RD UK\\
$^{9}$ ICRAR M468, University of Western Australia, 35 Stirling Hwy, Crawley WA 6009 Australia \\
$^{10}$ School of Physics \& Astronomy, University of St Andrews, North Haugh, St Andrews, KY16 9SS, Scotland \\
$^{11}$ Department of Physics and Astronomy, Macquarie University, NSW 2109, Australia \\
$^{12}$ Astronomy Centre, University of Sussex, Falmer, Brighton BN1 9QH }

\date{submitted to MNRAS, February 2015}

\pagerange{\pageref{firstpage}--\pageref{lastpage}} \pubyear{2015}

\maketitle

\label{firstpage}

\begin{abstract}
  We present the catalogue of blended galaxy spectra from the Galaxy And Mass
  Assembly (GAMA) survey.  These are cases where light from two galaxies are
  significantly detected in a single GAMA fibre.  Galaxy pairs identified from
  their blended spectrum fall into two principal classes: they are either strong
  lenses, a passive galaxy lensing an emission-line galaxy; or occulting
  galaxies, serendipitous overlaps of two galaxies, of any type. Blended spectra can 
  thus be used to reliably identify strong lenses for follow-up observations 
  (high resolution imaging) and occulting pairs, especially those that are a 
  late-type partly obscuring an early-type galaxy which are of interest for the 
  study of dust content of spiral and irregular galaxies. 
  The GAMA survey setup and its {\sc autoz} automated redshift determination
  were used to identify candidate blended galaxy spectra from the
  cross-correlation peaks. We identify \nsamp\ blended spectra with a 
  minimum velocity separation of 600 km/s, of which \nPEL\
  are lens pair candidates, \nELP\ emission-line-passive pairs, \nELEL\ are pairs of emission-line galaxies and and \nPP\ are pairs of galaxies with passive spectra.
  We have visually inspected the candidates in the Sloan Digital Sky Survey (SDSS) and Kilo Degree Survey (KiDS) images. Many blended objects are ellipticals with blue fuzz
  ({\em Ef} in our classification). These latter ``Ef" classifications are candidates for possible strong
  lenses, massive ellipticals with an emission-line galaxy in one or more
  lensed images.
  The GAMA lens and occulting galaxy candidate samples are
  similar in size to those identified in the entire SDSS. 
This blended spectrum sample stands as a testament of the power of this highly complete, second-largest 
  spectroscopic survey in existence and offers the possibility to expand e.g., strong gravitational lens surveys.
\end{abstract}


\section{\label{s:intro}Introduction}

Interstellar dust is still a dominant astrophysical unknown in cosmological distance estimates \citep{DETF, Holwerda08a,Holwerda14d} and models of how starlight is re-processed within a galaxy \citep[e.g.,][]{Baes10a, Bianchi11, Popescu11, de-Looze12b, Holwerda12a} because some 10-30\% of all the starlight is re-emitted by the dust in the far-infrared \citep{Popescu00}.
Interstellar dust can be found in two ways; by its emission or through the extinction of stellar light. 

Characterization of emission has made great strides with the {\em Spitzer} and {\em Herschel Space Observatories} \citep[e.g.,][]{Hinz09,Hinz12, Bendo12a,Bendo14,Smith10b,Baes10a,Xilouris12,Galametz12,Draine14,Verstappen13,Hughes14,Hughes14a}. A library of far-infrared and sub-mm images of nearby galaxies is currently being collated and more insight into the physics and distribution of interstellar dust in nearby galaxies can be expected with the great improvements in spectral coverage, sensitivity and spatial resolution.

Extinction measures of dust have some specific advantages over emission: they do not depend on the dust temperature, allowing for the detection of much colder dusty structures, and typically have the high resolution of the optical imaging observations. The single drawback is that one needs a known background light source.
In the case of the transparency of spiral galaxies, two techniques have just such a proven background source: background galaxies counts and occulting galaxy pairs. The technique that uses the number of background galaxies  \citep{Cuillandre01,Gonzalez98,Gonzalez03,Holwerda05,Holwerda05a,Holwerda05b,Holwerda05c,Holwerda05e,Holwerda05d,Holwerda07a,Holwerda07b,Holwerda12d} is nearing obsolescence as its inherent resolution and accuracy, limited by the intrinsic cosmic variance of background sources, are now surpassed by the accuracy and sensitivity of {\em Herschel Space Observatory} observations of dust surface density in nearby galaxies.

The occulting galaxies technique, however, has increased steadily in accuracy and usefulness, owing in a large part to the increasing sample sizes. Estimating dust extinction and mass from differential photometry in occulting pairs of galaxies was first proposed by \cite{kw92}. Their technique was then applied to all known pairs using ground-based optical images \citep{Andredakis92, Berlind97, kw99a, kw00a} and spectroscopy \citep{kw00b}.
Subsequently, some pairs were imaged with the {\em Hubble Space Telescope} \citep[HST][]{kw01a, kw01b, Elmegreen01,Holwerda09, Holwerda13b}. These initial results, however, were limited by sample sizes ($\sim15$ pairs).
More recently, new pairs were found in the SDSS spectroscopic catalogue \citep[86 pairs in][]{Holwerda07c} and through the GalaxyZOO project \citep{galaxyzoo}: 1993 pairs reported in \cite{Keel13}. This wealth of new pairs provided opportunities for follow-up with IFU observations \citep{Holwerda13a,Holwerda13b} and GALEX \citep{Keel14}. A greatly expanded occulting galaxy catalog improves accuracy as ``ideal pairs" -- an elliptical partially occulted by a late-type galaxy-- can be selected for follow-up. Ellipticals are the optimal background source as their light profile is smooth and very symmetric\footnote{The one exception is where we attempt to measure blue light attenuation. In this case, a spiral galaxy, which is brighter in the blue, is preferred.}.

Results from the occulting galaxy pairs include: (1) a mean extinction profile \citep{kw00a,kw00b,Holwerda07c}, (2) an indication that the dust may be fractal \citep{kw01a} and (3) the observation that the colour-extinction relation is grey,  i.e. there is little or no relation between the reddening of the stellar populations and total extinction.  The latter is due to the coarse physical sampling of ground-based observations. 
The Galactic Extinction Law returns as soon as the physical sampling of the overlap region resolves the molecular clouds in the foreground disk \citep[$<100$ pc][]{kw01a,kw01b, Elmegreen01,Holwerda09}.

A very reliable way to identify occulting galaxy pairs, i.e., purely serendipitous overlaps of galaxies is through blended spectra. In \cite{Holwerda07c}, we used the rejects from the Strong Lenses with ACS Survey \citep[SLACS][]{Bolton04,slacs1}, a highly successful search for strong lenses, confirmed with HST \citep{slacs2,slacs3,slacs4,slacs5,slacs6,slacs7,slacs8}, with spectroscopic selection extended now to the BOSS survey \citep{Brownstein12, Bolton12a}. Both types of blended spectral sources have two things in common: very close association on the sky (within a SDSS spectroscopic fibre of 3'' diameter) and clear spectroscopic signal from both galaxies at distinct redshifts. 

In this paper, we present the blended spectra catalogue based on the Galaxy And Mass Assembly (GAMA) survey \citep{Driver09, Driver11, Baldry10} as candidates for either strong lensing follow-up or occulting galaxy analysis (e.g., HST imaging or spectroscopy). The GAMA data is an improvement over SDSS as the target galaxies can be fainter, the aperture is smaller (i.e., a closer overlap of the galaxies) and the {\sc autoz} detection algorithm is a marked improvement on the SDSS detections.
The paper is organized as follows: 
\S \ref{s:gama} briefly introduces the GAMA survey, 
\S \ref{s:selection} the redshift determination and selection of blended spectra, 
\S \ref{s:vis} presents the visual classifications of the blended objects, 
\S \ref{s:cat} presents the blended spectra catalogue and we discuss the pair classification and their possible future uses in 
\S \ref{s:disc}.

\section{Galaxy And Mass Assembly (GAMA) survey}
\label{s:gama}

The GAMA survey has obtained over 250\,000 galaxy redshifts 
selected to $r < 19.8$\,mag over 290 deg$^2$ of sky 
\citep{Driver09,Driver11,Baldry10,Liske14}.  
At the heart of this survey is
the redshift survey with the upgraded 2dF spectrograph AAOmega \citep{Sharp06,Saunders06}
on the Anglo-Australian Telescope.  The GAMA survey extends over three
equatorial survey regions of 60 deg$^2$ each (called G09, G12 and G15) and two
Southern regions of similar area (G02, G23). See \cite{Baldry10} for a
detailed description of the GAMA input catalogueue for the equatorial regions.

The redshift survey in combination with a wealth of imaging data has led to
many science results already. We use for this work the GAMA~II redshifts
\citep{Liske14}, which were obtained using a robust cross-correlation method
for spectra with and without strong emission lines \citep{Baldry14}.

\section{Selection of Blended Spectra}
\label{s:selection}

Galaxy redshifts were initially determined by a supervised fit \citep{Liske14}
but a recent upgrade to the GAMA survey pipeline includes a fully-automated
template-based redshift determination \citep[{\sc autoz},][]{Baldry14}. In
certain cases, the fits for different templates resulted in two high-fidelity, 
but different redshifts; these are the candidate blended objects of interest to us here.

The {\sc autoz} code obtains cross-correlation redshifts against stellar and
galaxy templates with varying strength of emission and absorption line
features. The height and position of the first four peaks of normalized
cross-correlation functions are obtained. 
These are called $r_x$, $r_{x,2}$,
$r_{x,3}$ and $r_{x,4}$ each with a corresponding redshift and template
number, with the peaks separated by at least 600\,km/s.  High values of $r_x$
and $r_{x,2}$, particularly relative to $r_{x,3}$ and $r_{x,4}$, can then be
used to select candidate blended spectra. 

The {\sc autoz} algorithm marks a significant improvement in the
identification of blended spectra over that which could be obtained from
GAMA~I redshifts. In the initial redshift campaign, ``redshifters'' -- the GAMA
team members identifying the redshift with {\sc runz} -- were focused on
attaining a reliable redshift for single objects. In such an approach, only
those spectra with wildly different redshifts by two redshifters or spectra
remarked upon during visual inspection would be selected. With {\sc autoz},
blended spectra are identified as different redshifts using normalized
cross-correlation functions, a much more objective and complete approach.

\subsection{AUTOZ selection}

\begin{figure}
\begin{center}
\includegraphics[width=0.45\textwidth]{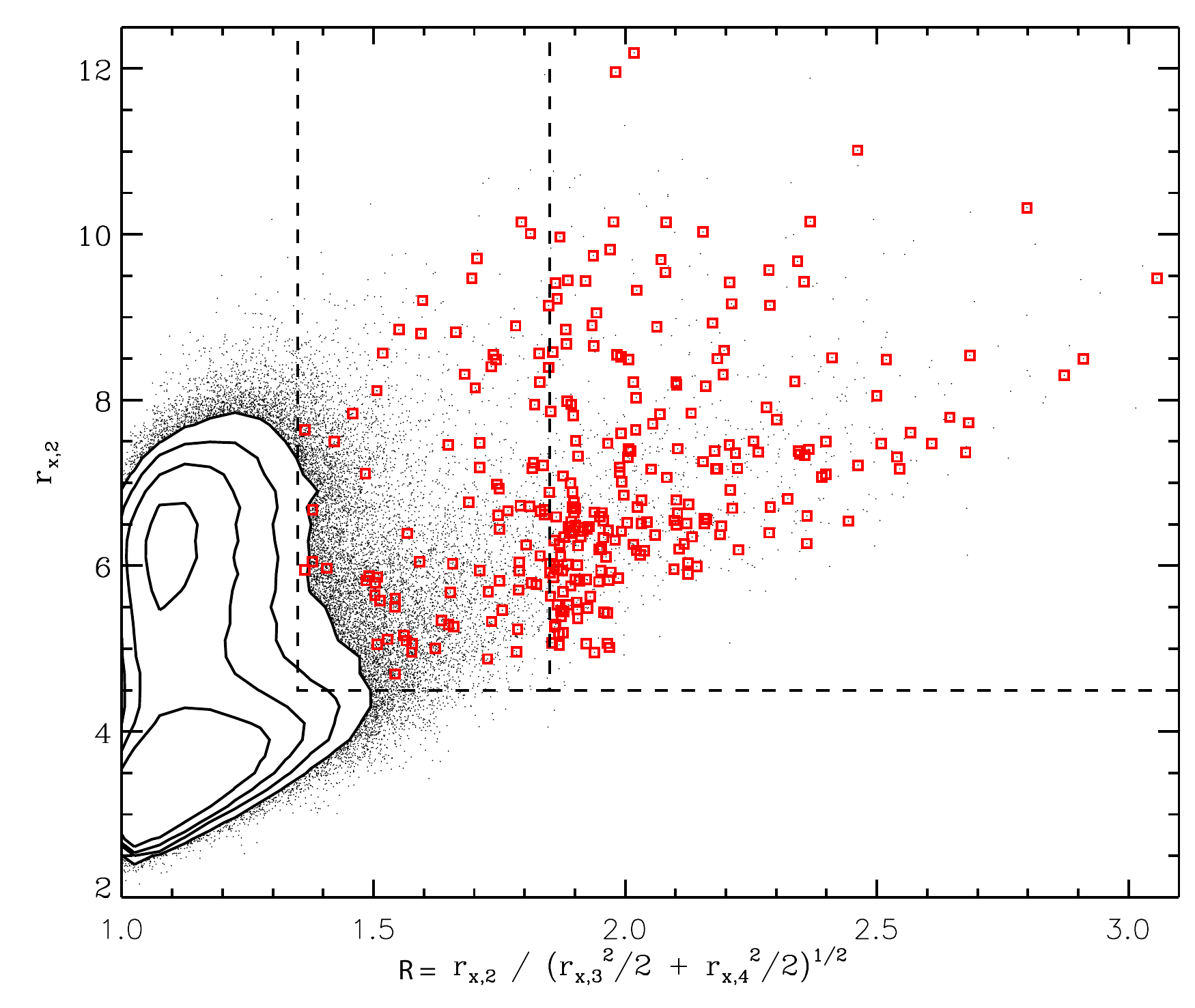}
\caption{Distribution in selection parameters: for all GAMA spectra
  shown with contours and black points, and selected candidates
  shown with red squares. Note the all GAMA sample includes star-galaxy
  blends. The main selection criteria was ${\cal R} > 1.85$ and this results
  in $r_{x,2} > 4.5$ by default. Candidates from an earlier version of the
  code have ${\cal R}$ between 1.35 and 1.85. These boundaries are shown
  with dashed lines.}
\label{f:double-z}
\end{center}
\end{figure}

\begin{figure}
\begin{center}
\includegraphics[width=0.45\textwidth]{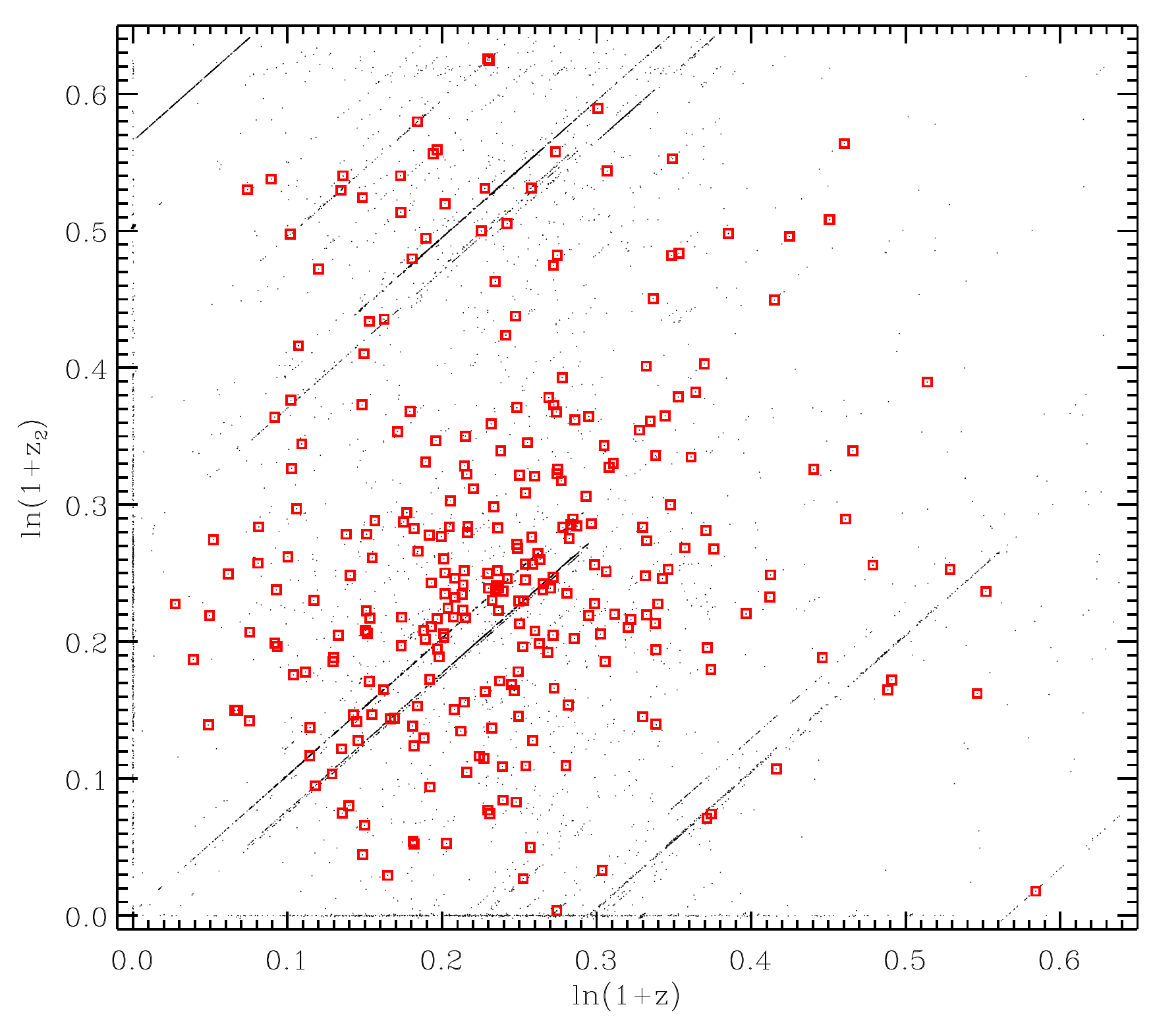}
\caption{Distribution in $z_2$ versus $z$: for all GAMA spectra
  with ${\cal R} > 1.35$ and $r_{x,2} > 4.5$ shown with black points, 
  and for the selected candidates ( ${\cal R} > 1.85$) shown with red squares. Most candidates
  lie away from the alias lines that are appear parallel when plotting 
  the logarithm of one plus redshift on each axis.}
\label{f:z-z-plot}
\end{center}
\end{figure}

Double redshift selection using the {\sc autoz} approach was to require that
two different redshifts had high cross-correlation peaks 
($r_x$ and $r_{x,2}$), while the next two redshifts had significantly 
lower peak values. 
In order to address this, we defined the ratio between the second redshift 
peak value and subsequent, third and fourth, redshift peaks to be:
\begin{equation}
{\cal R} = r_{x,2} / \sqrt{(r_{x,3}^2 + r_{x,4}^2)} 
\end{equation}
To select the double-z candidates, we required ${\cal R} > 1.85$  (to avoid aliasing and a clean selection of real blends, see Figures \ref{f:double-z} and \ref{f:z-z-plot}) and
for the first two redshifts to be from galaxy spectral templates.  The
galaxy spectral templates used in {\sc autoz} were from SDSS.  An
early version of the code used the SDSS DR2 templates, while a later
version used templates derived from the \citet{Bolton12a} galaxy
eigenspectra.  These later templates were numbered 40-47 in order of
increasing emission-line strength.  To broadly classify the templates,
we select templates 40-42 as `passive galaxies' (PG) and 43-47 as
`emission-line galaxies' (ELG).

Initially, we selected candidates with ${\cal R}>1.85$ and $r_{x,2}>
5.5$, with these values determined using the early version of the
code.  After the code was updated, the value of ${\cal R}$ changed as
a result of the new templates and a modest increase in the allowed
redshift range from 0.8 to 0.9. Using the new values, we selected
candidates with ${\cal R}>1.85$ and with no restriction on
$r_{x,2}$. Old candidates were retained subject to a couple of
criteria: the new value of ${\cal R}$ was still greater than 1.35, and
$(1+z)/(1+z_2)$ was not near a problematic cross-correlation alias.
Aliases can result from the matching of different emission lines
in the templates to a strong line in the data.  All candidates, old
and new, near the alias of $(1+z)/(1+z_2) = 1.343\pm0.002$ ($\sim
5007/3727$) or the inverse were removed from the
sample. Figure~\ref{f:double-z} shows the distribution of candidates
in $r_{x,2}$ versus ${\cal R}$, i.e.\ the selection
parameters. Figure~\ref{f:z-z-plot} shows where the aliases lie, in
$z_2$ versus $z$, with respect to the candidates.

The selection resulted in \nsamp\ galaxy pair candidates (from 299
blended spectra -- some source locations were observed more than
once).  Depending on which template matched best for both redshifts in
the blend, we classified the type of blends as follows: two passive-template
galaxies (PG+PG), two emission-line templates (ELG+ELG), a passive
template at low redshift and emission line template at higher redshift
(PG+ELG), or vice versa (ELG+PG). Table \ref{t:spectype} summarizes
the resulting classifications. The {\sc autoz} results for our \nsamp\
blended objects are listed in Table \ref{t:cat}.

\begin{table}
\caption{The numbers of different blended spectra identified in GAMA using the {\sc autoz} algorithm.}
\begin{center}
\begin{tabular}{cr}
Pair Type & Number \\ \hline \hline
ELG+ELG   &       78 \\
ELG+PG    &       71 \\
PG+ELG    &      104 \\
PG+PG     &       27 \\ \hline
Total   & 280 \\
\end{tabular}
\end{center}
\label{t:spectype}
\end{table}%

\section{Visual Classification}
\label{s:vis}

We visually classified all the \nsamp\ galaxy pairs identified by {\sc autoz} using the SDSS image viewer and GAMA cutouts in the case of the Southern fields. We classify whether the object appears as a single galaxy (either S or E, where S can mean a spiral galaxy or an irregular one i.e., late-type, except in a few clear cases and E an early type), an occulting pair or a disturbed ongoing merger (M). We sub-classify the occulting pair similar to the \cite{Keel13} classification (Table \ref{t:classification}) but given the limited image resolution the classification is essentially S-E, S-S or E-E. We introduce a sub-classification for single early types with some blue fuzzy hue on one side of the SDSS {\em gri} composite galaxy images (Ef). These latter could be lensing ellipticals or simply occulting pairs with a very small irregular galaxy in the fore- or background.  

\begin{figure}
\begin{center}
\includegraphics[width=0.5\textwidth]{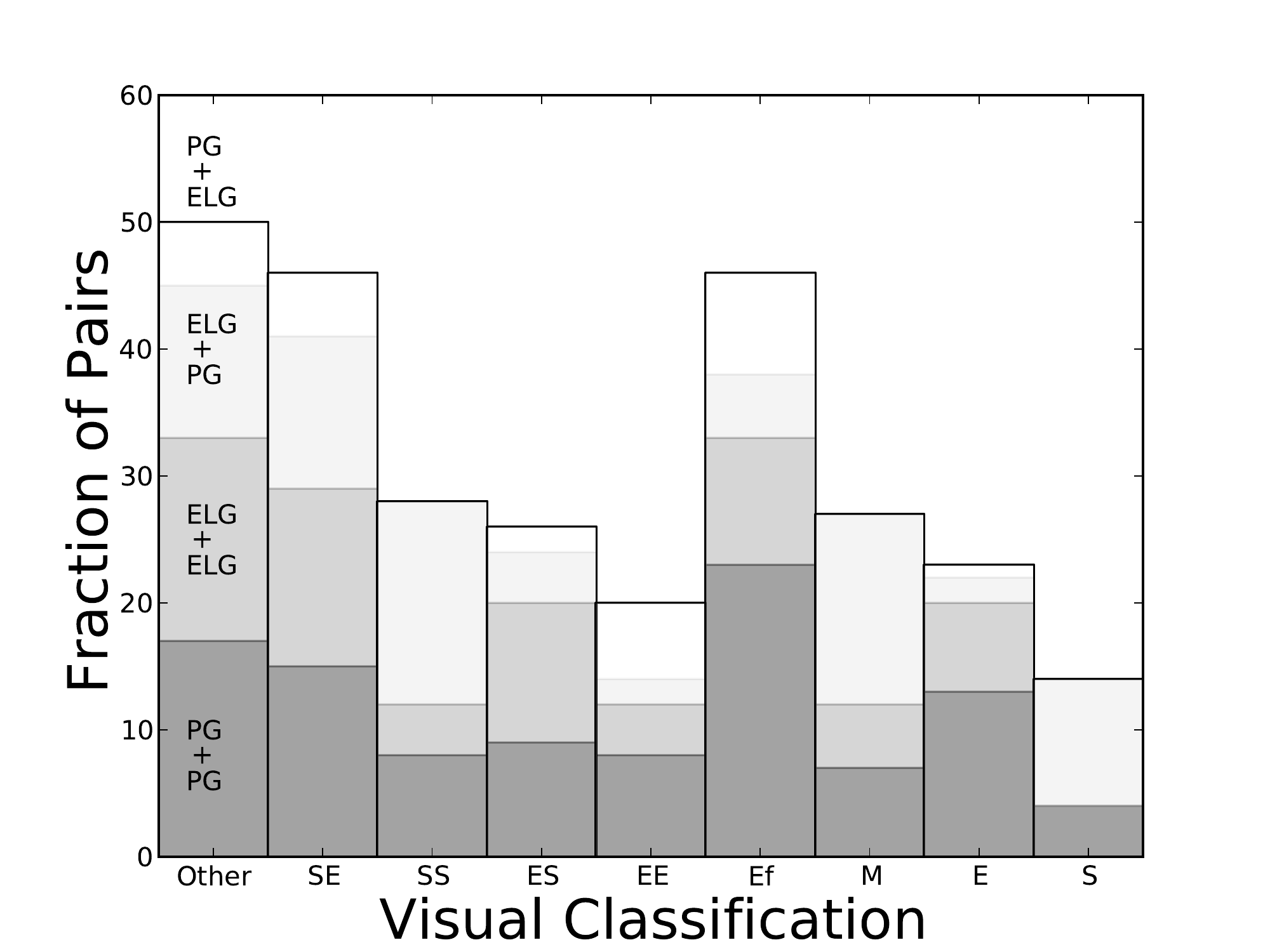}
\caption{The  distribution of visual classifications for all the blended spectra broken down into all four possible template combinations (PG=passive template, ELG=emission line template, PG+ELG is a passive in front of an emission line pair). No clear correlation between the SDSS visual classifications and the template ones is evident. The "other" category are objects without SDSS imaging or a reasonable classification. }
\label{f:hist:vc}
\end{center}
\end{figure}

\begin{figure}
\begin{center}
\includegraphics[width=0.45\textwidth]{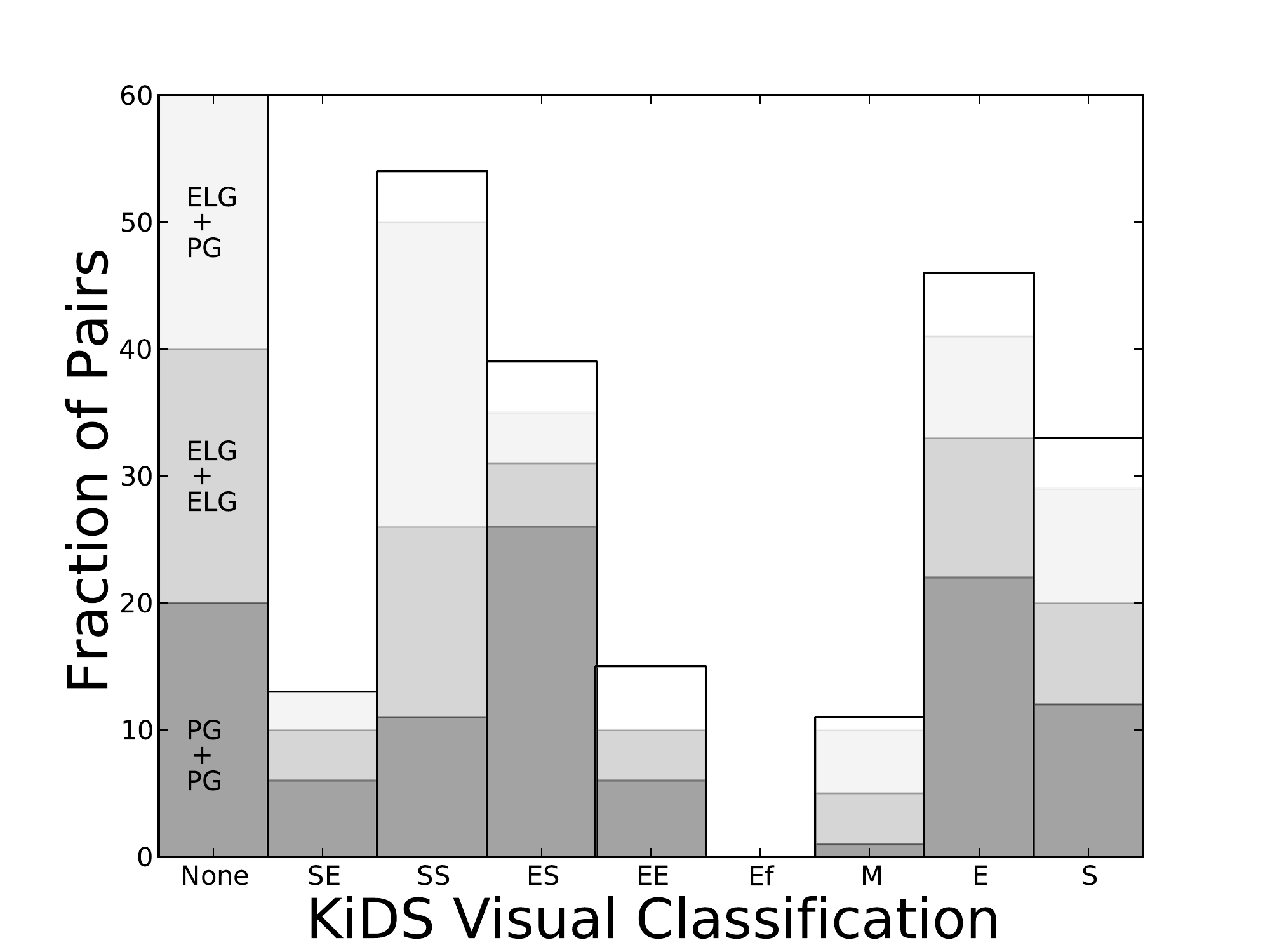}
\caption{The distribution of visual classifications of the KiDS {\em sdss-i} filter for all the blended spectra broken down into all four possible template combinations (PG=passive template, ELG=emission line template, PG+ELG is a passive in front of an emission line pair). The lack of a correlation between visual and spectral classification persist with the singe-filter classifications.}
\label{f:hist:kvc}
\end{center}
\end{figure}

The classifications are listed in Table \ref{t:classification}. Figure \ref{f:hist:vc} shows the distribution of visual classifications of the blended spectrum objects. One would expect a reasonable correlation between the spectral typing (e.g., passive vs. emission-line) with the visual classifications, e.g., passive spectra dominating those pairs with an E type galaxy in the foreground. 
{
The relation between visual classification and spectral classification is tenuous. Visual classification based on color images can be both powerful but also misleading. Blue galaxies tend to be classified as late-types even if their profile is actually that of a spheroid. 

With this in mind, two of us (BWH and AM) re-classified {\em sdss-i} postage stamps from the KiLO Degree Survey (KiDS) survey \citep{kids}. Figure \ref{f:hist:kvc} shows the distribution of these visual classifications. Because these are single-filter, the classification {\em Bf} is impossible. The new visual classifications remains poorly correlated with the spectral classification.

In our opinion, both the SDSS color-images or the deeper and higher resolution KiDS single-filter images are still too low-resolution to unambiguously disentangle and visually classify these objects. These objects are inherently blended ones. Even with another improvement in spatial resolution (i.e., HST imaging), visual classifications will remain subjective --although it is encouraging that BWH and AM agreed on the visual classifications. And it remains difficult to ascertain which object is in the foreground in a visual classification.

}


\begin{table}
\caption{The \protect\cite{Keel13} classification of the occulting galaxy pairs.}
\begin{center}
\begin{tabular}{ll}
Classification 	& Description \\
\hline
\hline
F			& spirals seen nearly face-on in front of an \\
			& elliptical or S0 background system. \\
Q 			& the background galaxy is nearly edge-on and \\
			& is projected nearly radial. \\
$\Phi$		& the spiral is seen essentially edge-on, \\
			& at least partially backlit by a smooth galaxy. \\
X			& two edge-on disk galaxies \\
SE			& Spiral in front of an Elliptical, \\
			& not in one of the above categories. \\
S			& spiral/spiral overlaps. \\
B			& the background galaxy has much smaller \\
			& angular size than the foreground disk. \\
E			& pairs containing only elliptical or S0 galaxies. \\
\hline
\end{tabular}
\end{center}
\label{t:classification}
\end{table}%

\begin{table*}
\caption{\label{t:cat} The complete catalogue of blended spectra in the GAMA survey. T1 and T2 refer to the template numbers for the first and second peaks. 
The full catalog is available in the appendix. }
\begin{center}
\begin{tabular}{r r r r r r r r r c c c}
Field & GAMA-id & RA & DEC & $z$ & T1 & $r_x$ & $z_2$ & T2 & $r_{x,2}$ & Spec.\ Type & Vis.\ Type \\
\hline
\hline
 G09&  196060& 129.01621&  -0.69336& 0.293&40&  8.7& 0.051&46&  5.6& ELG+PG & SE    \\
 G09&  197073& 133.78179&  -0.74790& 0.270&40& 10.8& 0.268&44&  6.4& ELG+PG & EE    \\
 G09&  198082& 138.28150&  -0.66673& 0.163&40& 11.1& 0.321&47& 10.2& PG+ELG & ES    \\
 G09&  202448& 129.69546&  -0.38179& 0.418&40&  9.0& 0.738&45&  5.0& PG+ELG & SE    \\
 G09&  204140& 136.63883&  -0.35203& 0.282&40&  9.4& 0.449&47&  8.1& PG+ELG & Ef    \\
 G09&  209222& 132.36771&   0.16360& 0.128&40& 10.3& 0.603&47&  6.7& PG+ELG & E     \\
 G09&  209263& 132.50596&   0.04250& 0.310&42&  6.5& 0.270&46&  5.3& ELG+PG & Ef    \\
 G09&  209295& 132.61013&   0.11972& 0.313&40& 11.2& 0.608&47&  7.8& PG+ELG & Ef    \\
\dots & \dots & \dots & \dots & \dots & \dots & \dots & \dots & \dots & \dots & \dots & \dots  \\
\hline 
\end{tabular}
\end{center}
\end{table*}%

\section{Catalog}
\label{s:cat}

\begin{figure}
\begin{center}
\includegraphics[width=0.5\textwidth]{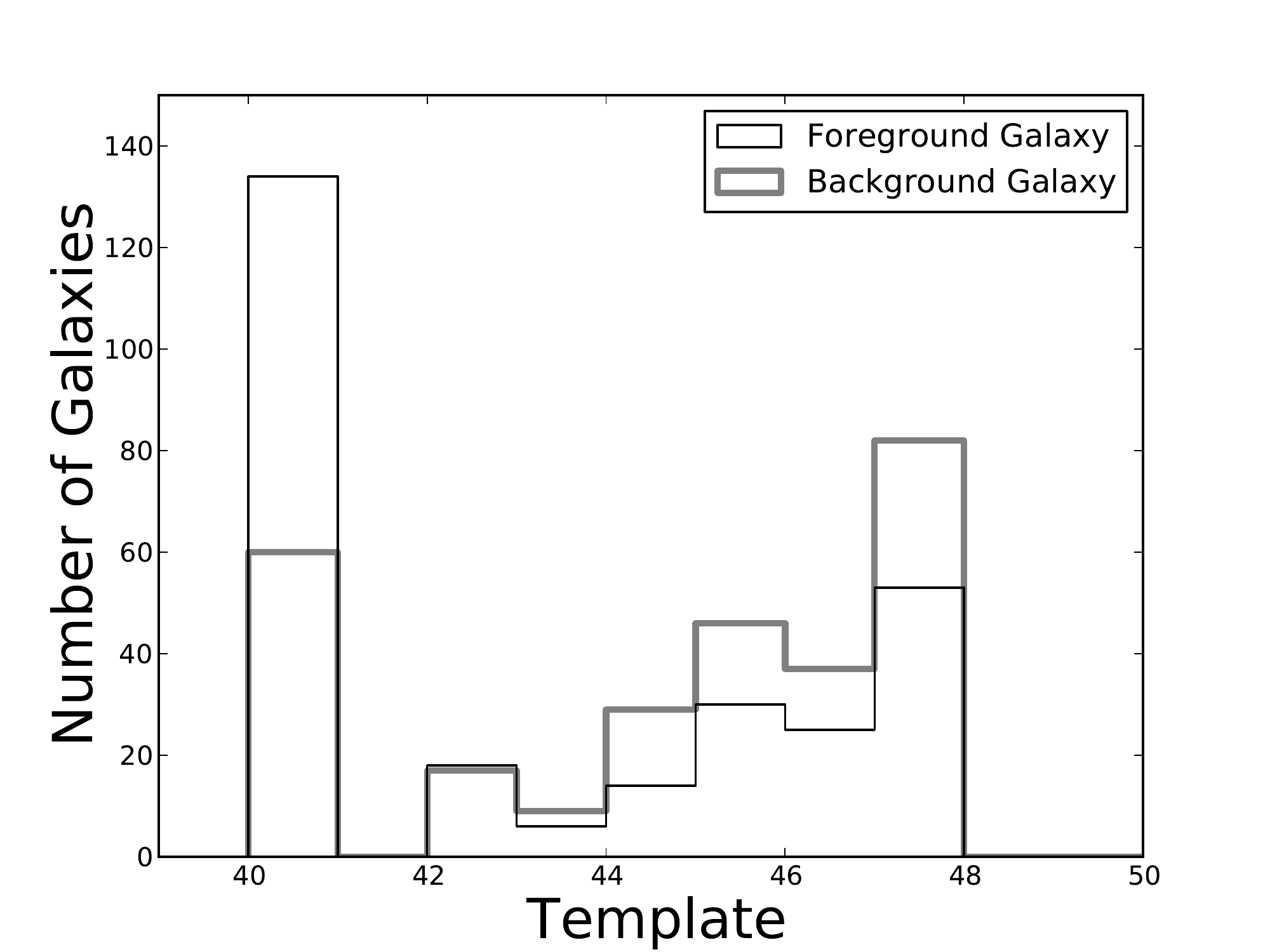}
\caption{The distribution of GAMA spectroscopic best template for the blended spectra catalogue. 
The templates are numbered from 40--47 in order of increasing emission-line strength.
To broadly classify the templates, we select templates 40-42 as ``passive galaxies'' (PG) and 43-47 as ``emission-line galaxies'' (ELG).}
\label{f:temphist}
\end{center}
\end{figure}

We have classified the blended spectra by the best-fit templates (PG=passive template, ELG=emission line template) denoting them with Foreground+Background best fit template.
Out of \nsamp\ galaxies, we identify: 
\nPEL\ lens candidates, passive galaxies with emission-line galaxies at higher redshifts (PG+ELG); 
\nELP\ ideal occulting galaxy pairs, emission-line galaxies in front of a passive galaxies (ELG+PG); 
\nELEL\ occulting pairs with both foreground and background galaxies showing strong emission lines, ELG+ELG; 
and \nPP\ double passive occulting pairs (PG+PG), with both galaxies having passive template fits 
(Table \ref{t:spectype}). Figure \ref{f:temphist} shows the distribution of foreground and
background galaxy best-fit templates. We characterize PG+ELG pairs as possible lensing pairs as this is how
the SLACS survey found the majority of their strong gravitational lensing pairs, confirmed with HST imaging \citep{slacs2,slacs3,slacs4,slacs5,slacs6,slacs7,slacs8}.
%
There is a preference for template 40 in the case of foreground galaxies. We interpret this as a selection effect: it is easier to identify anomalous emission lines on top of a passive spectrum (template 40 has the weakest emission lines). 
Apart from the preference for template 40 for foreground objects, the distribution is relatively similar. The  {\sc autoz} classification is not particularly biased against either combination of spectra in a blend.

\begin{figure}
\begin{center}
\includegraphics[width=0.5\textwidth]{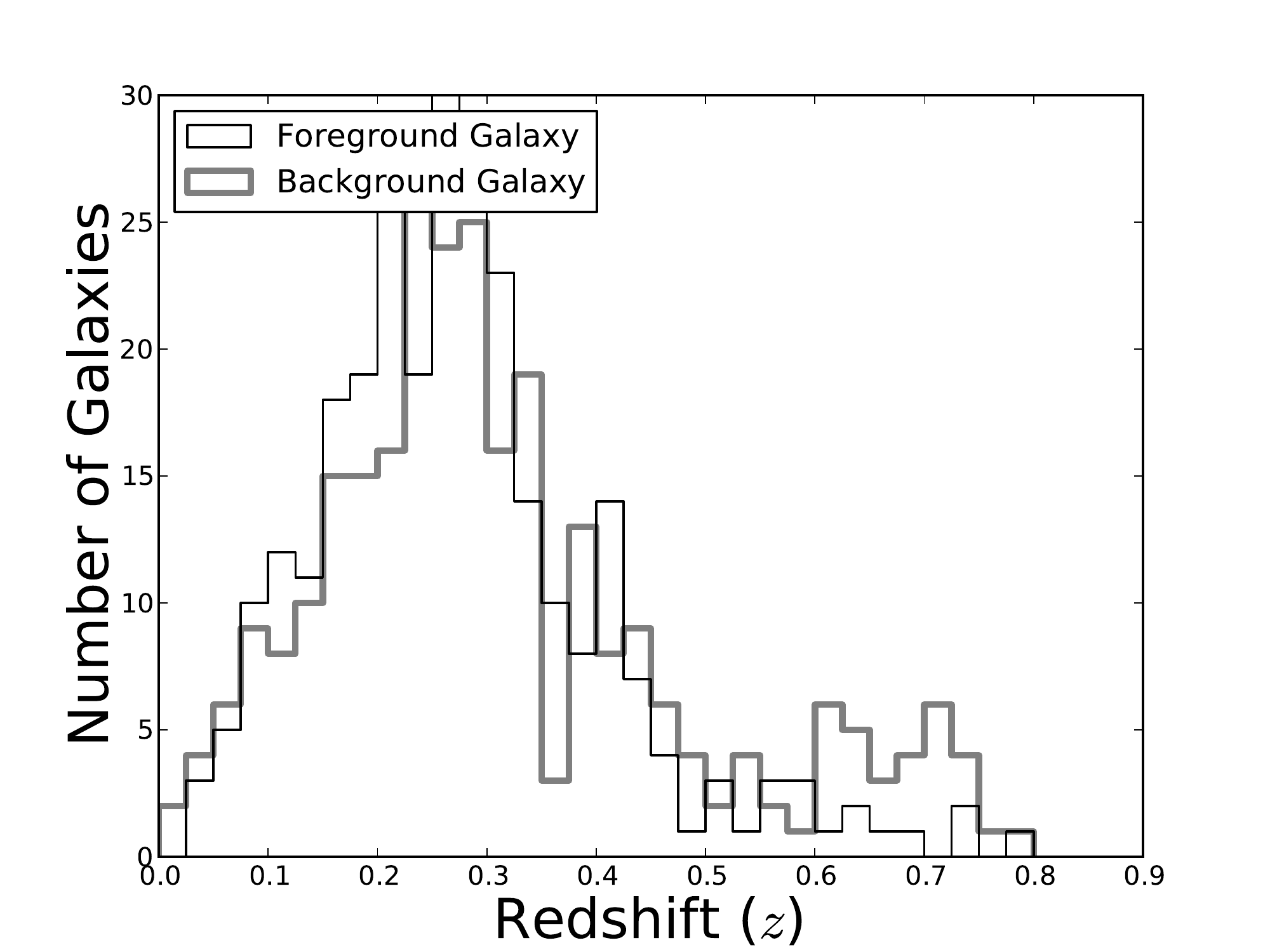}
\caption{The distribution of GAMA spectroscopic redshifts for the blended spectra catalogue. {\sc AUTOZ} excludes redshift candidates within 600 km/s of another redshift by design. }
\label{f:zhist}
\end{center}
\end{figure}

\begin{figure}
\begin{center}
\includegraphics[width=0.5\textwidth]{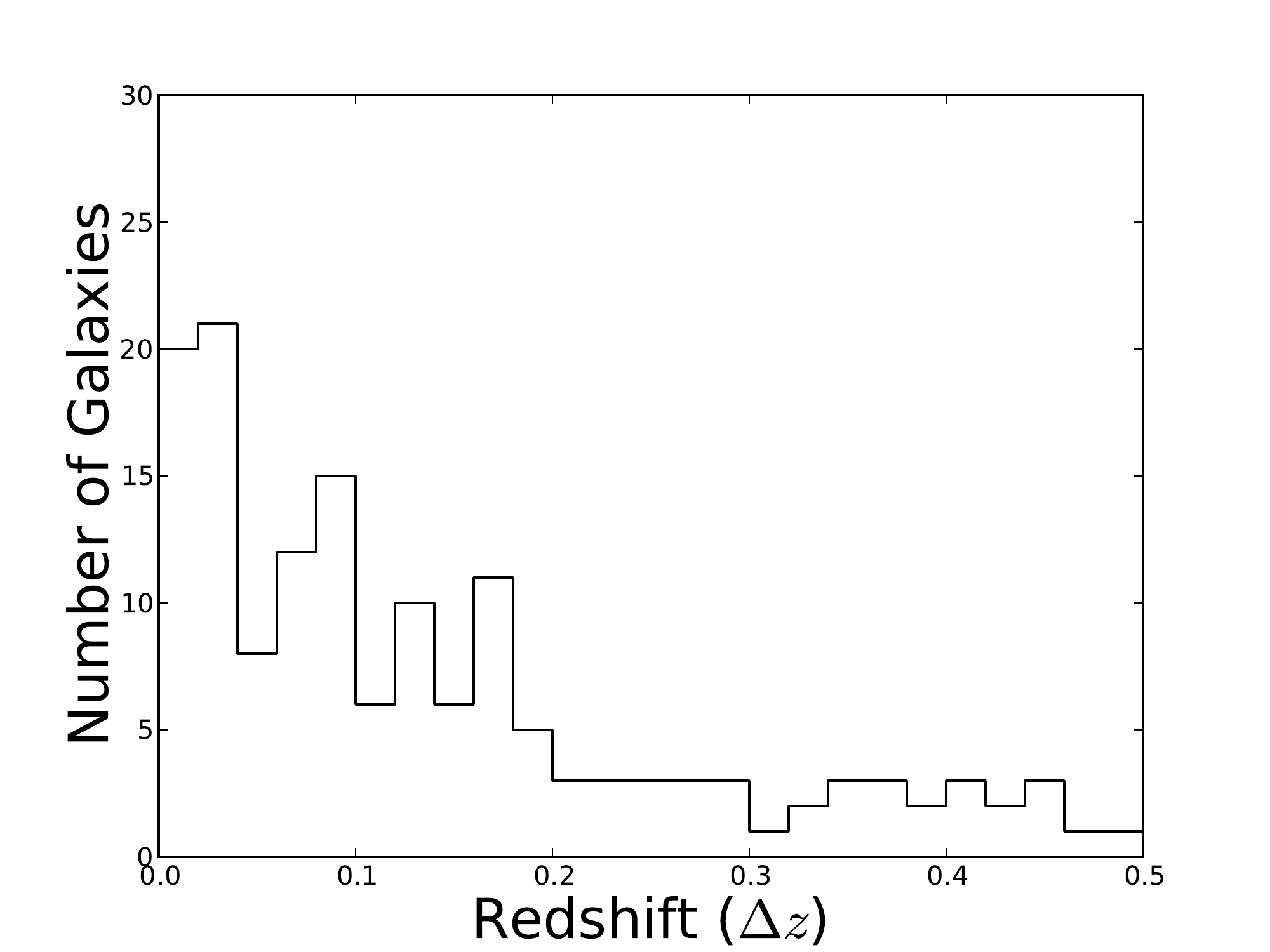}
\caption{The difference in redshift between the foreground and background galaxies in blended spectra catalogue.} 
\label{f:deltaz}
\end{center}
\end{figure}

Figure \ref{f:zhist} shows the distribution of redshifts for both the foreground and background galaxies in the blended spectra. 
Background galaxy redshifts peak around $z=0.3$ and foreground galaxies a little below that. Similar to the redshift completeness 
of the GAMA survey \citep{Baldry14,Liske14}, the sample is complete for $z<0.4$ regardless of template but objects can still be detected out to $z<0.8$, 
beyond which the {\sc autoz} are limited because of a lack of information in the GAMA spectra.
The pair members are typically well  separated in redshift ($\Delta z > 600 km/s$). 
%
Figure \ref{f:deltaz} shows the redshift difference for the blended spectra, making these ideal pairs for either lensing studies or as occulting galaxies.

To classify the pairs into strong lens candidates (PG+ELG) and ELG+PG, ELG+ELG, and PG+PG occulting pairs, we employ the best template fits. 
Lenses (PG+ELG) are difficult to verify from ground-based imaging, but \cite{Arneson12} argue that spectroscopic selection of lenses are both complete and relatively unbiased within the Einstein ring. Therefore the list of lenses presented here, especially those in the G23 field (not covered by SDSS), are new candidates for possible HST follow-up.
Figure \ref{f:ocgexamples} shows some random examples of ``ideal'' occulting pairs (ELG+PG).
In \cite{Holwerda07c}, we found that 86 out of 101 candidate from SLACS were usable occulting pairs. Figure \ref{f:ocgexamples} shows that indeed most of the spectroscopically-identified occulting pairs have a good geometry to extract, in principle, the transmissivity of the foreground galaxies. These new pairs will be of use to model the transmission of the foreground galaxy with a very low impact parameter (almost perfectly aligned galaxies).
Alternate occulting galaxy pairs are the ELG+ELG type, which can be used to extract transmission though the foreground galaxy in the bluer wavelengths. For example, \cite{Keel14} use such spiral-spiral pairs to infer the extinction law in the ultraviolet. This can only be done with a UV-bright spiral as the background galaxy. Intrinsic asymmetry in spiral structure of both pair members introduces uncertainty in the transmission/opacity measurement, but does not introduce a bias. However, irregular galaxies cannot be used as background illuminators.
The ELG+ELG occulters are therefore a useful sub-sample of the occulting pairs. 
Certainly, one is a clear ELG+ELG pair but many other include an irregular as well.
Lastly, we have PG+PG pairs, where both galaxies lack emission lines. These may be lenses still, but are unlikely to attract follow-up attention. As occulting galaxies they are not likely to reveal much new information about the dusty ISM in early-types.

\begin{figure*}
\begin{center}
\includegraphics[width=0.49\textwidth]{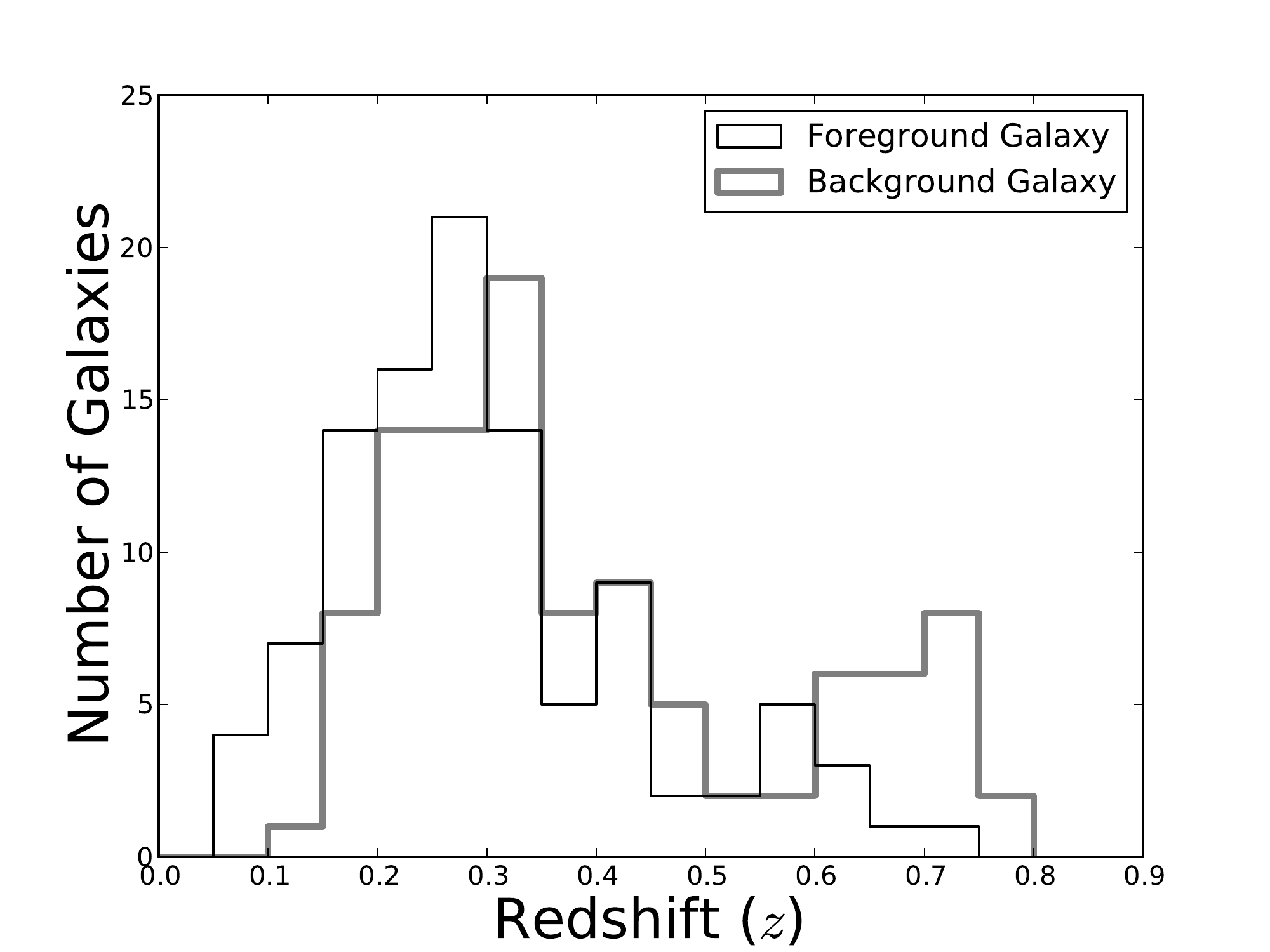}
\includegraphics[width=0.49\textwidth]{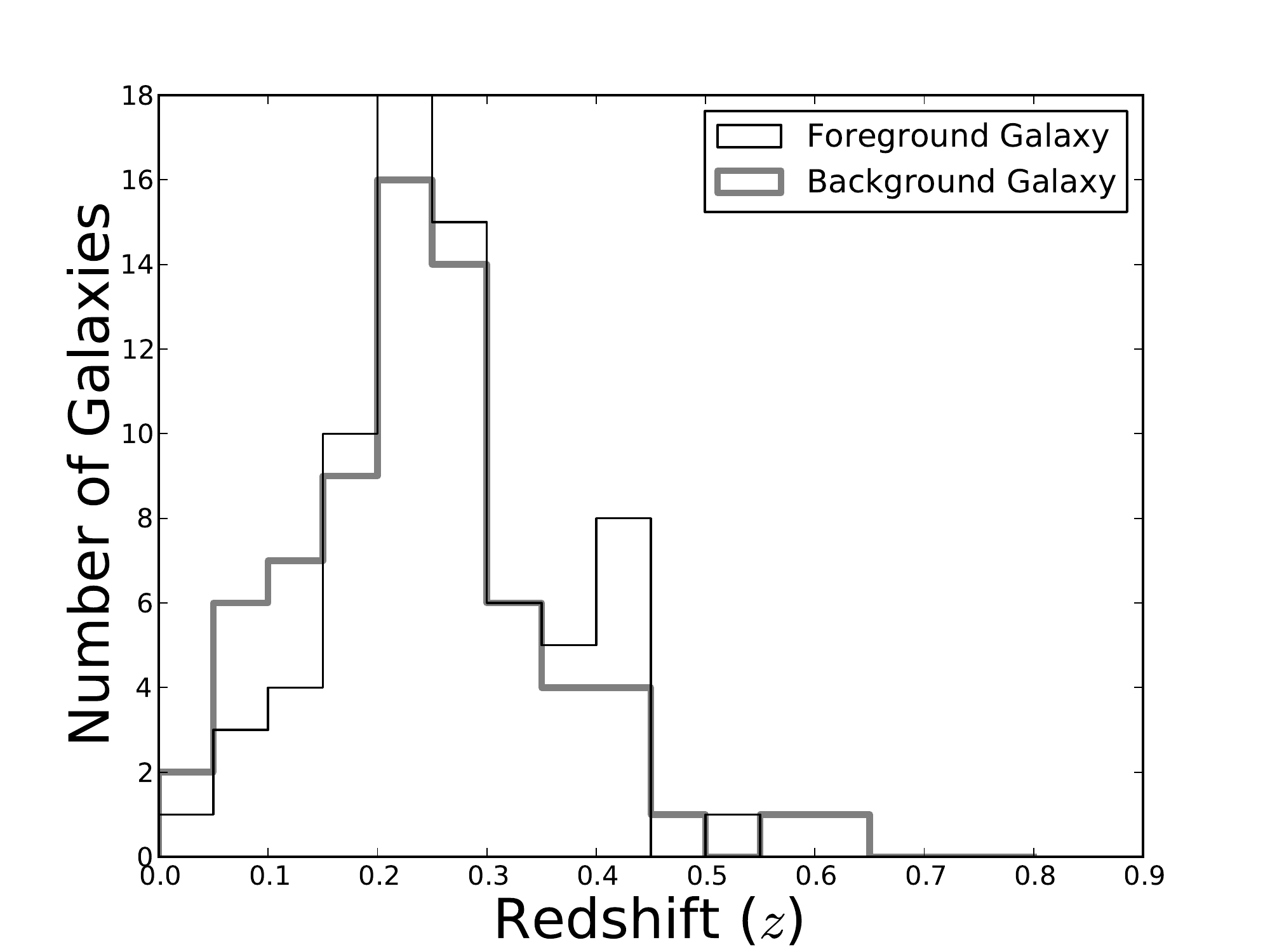}
\caption{The distribution of foreground galaxy redshift and background galaxy redshift for 
the passive foreground with emission-line background templates (PG+ELG, left panel) and 
the emission-line foreground template with a passive template for the background objects (ELG+PG, right).}
\label{f:pairs:z}
\end{center}
\end{figure*}


\begin{figure*}
\begin{center}
\includegraphics[width=0.49\textwidth]{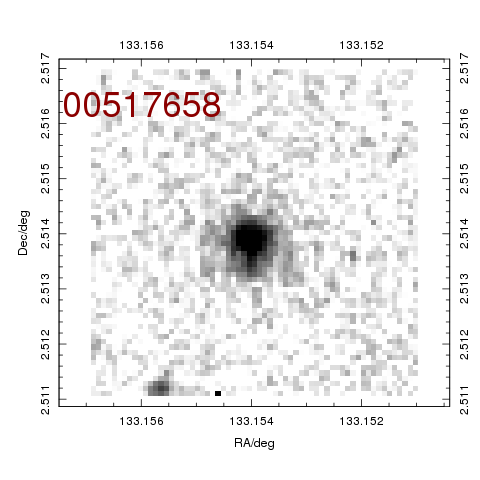}
\includegraphics[width=0.49\textwidth]{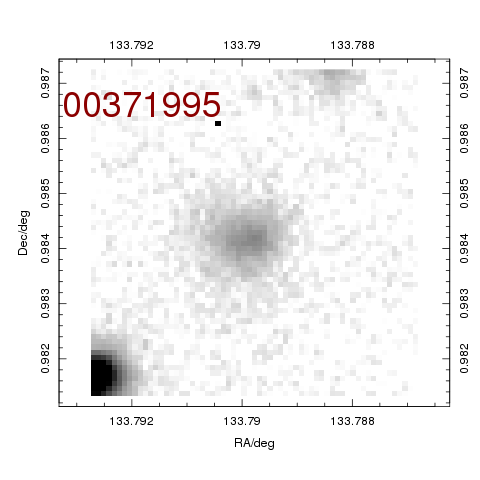}\\
\includegraphics[width=0.49\textwidth]{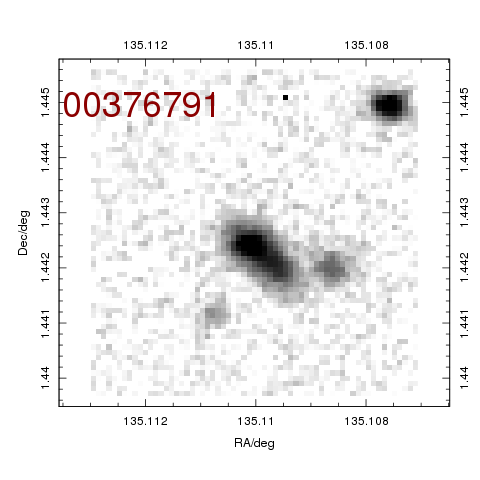}
\includegraphics[width=0.49\textwidth]{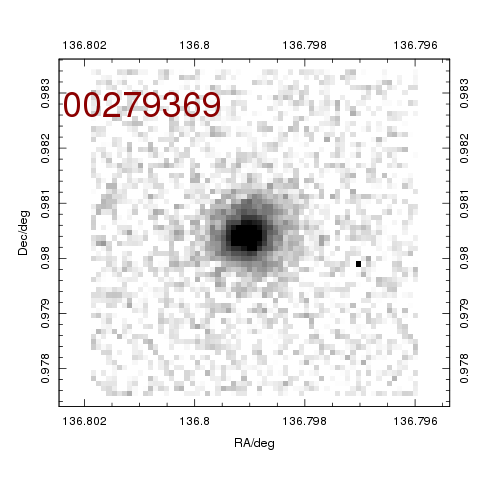}\\
\caption{Examples of the ELG+PG pairs. SDSS-i image cutouts retrieved using \url{http://ict.icrar.org/cutout/}. Standard cutout size is 20". GAMA id numbers are in red.  }
\label{f:ocgexamples}
\end{center}
\end{figure*}

\begin{figure*}
\addtocounter{figure}{-1}
\begin{center}
\includegraphics[width=0.49\textwidth]{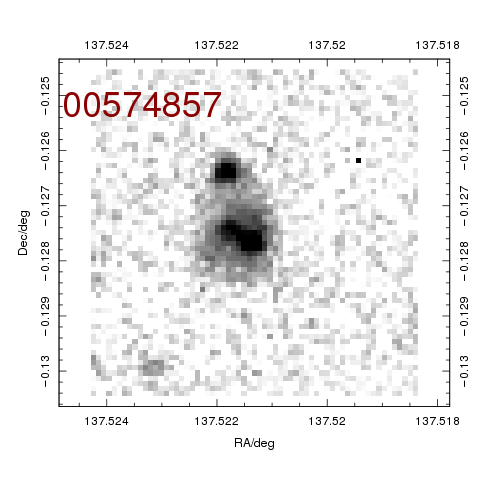}
\includegraphics[width=0.49\textwidth]{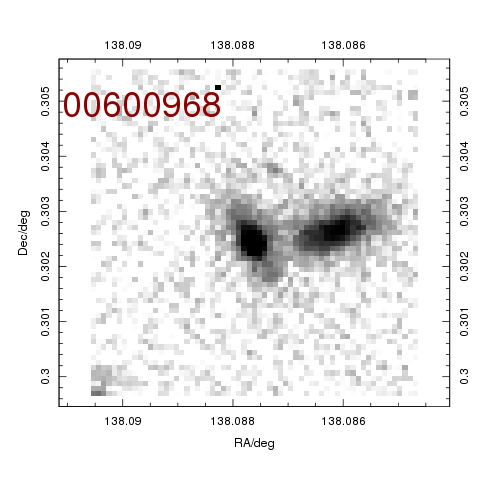}\\
\includegraphics[width=0.49\textwidth]{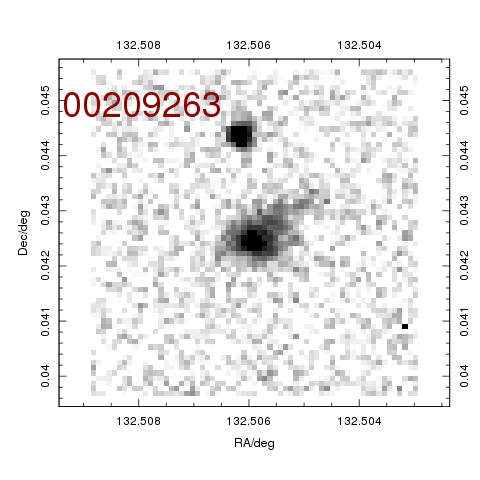}
\includegraphics[width=0.49\textwidth]{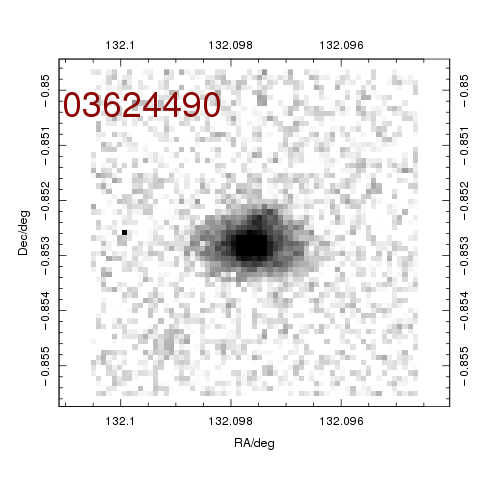}\\
\caption{-- {\em continued}}
\end{center}
\end{figure*}

\begin{figure*}
\begin{center}
\includegraphics[width=0.49\textwidth]{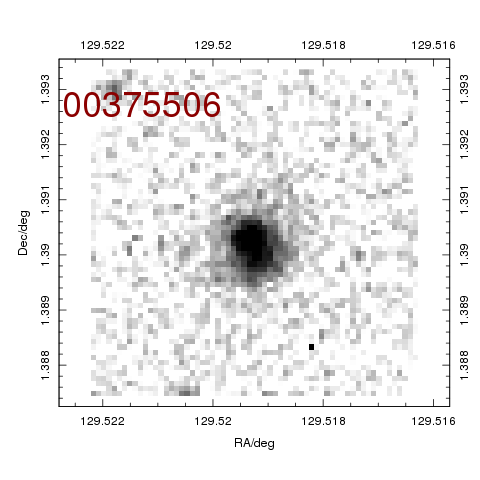}
\includegraphics[width=0.49\textwidth]{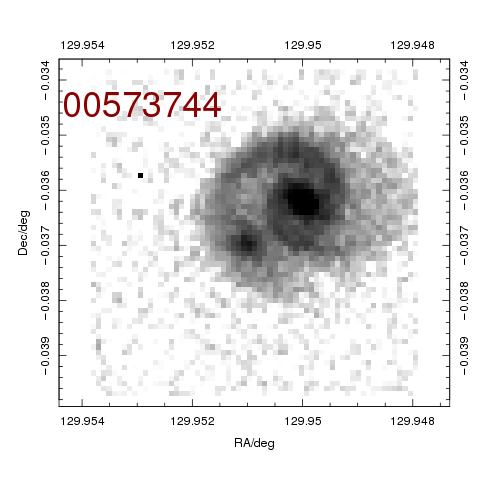}\\
\includegraphics[width=0.49\textwidth]{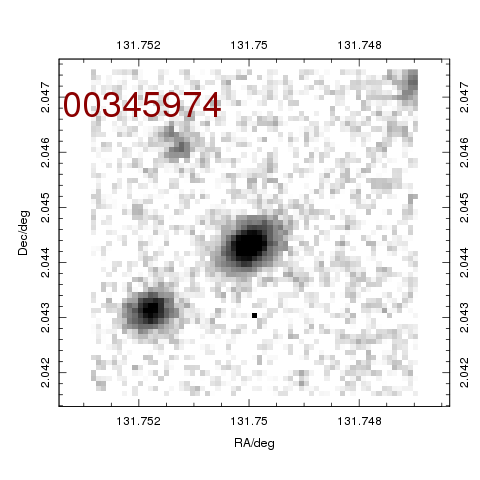}
\includegraphics[width=0.49\textwidth]{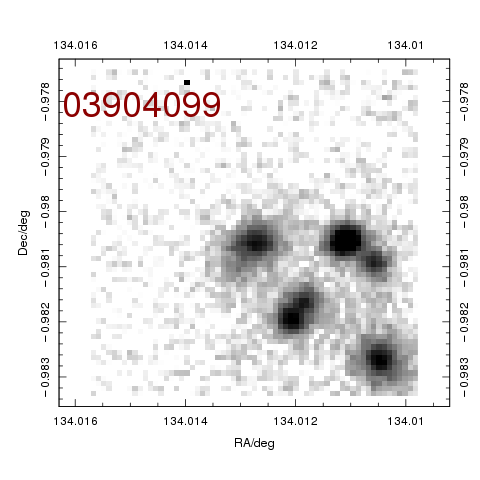}\\
\caption{Examples of the ELG+ELG pairs. SDSS-i image cutouts retrieved using \url{http://ict.icrar.org/cutout/}.  Standard cutout size is 20". GAMA id numbers are in red. Emission line galaxies include irregulars which limit the use of this class of objects in follow-up analysis. }
\label{f:uvocgexamples}
\end{center}
\end{figure*}

\begin{figure*}
\begin{center}
\addtocounter{figure}{-1}
\includegraphics[width=0.49\textwidth]{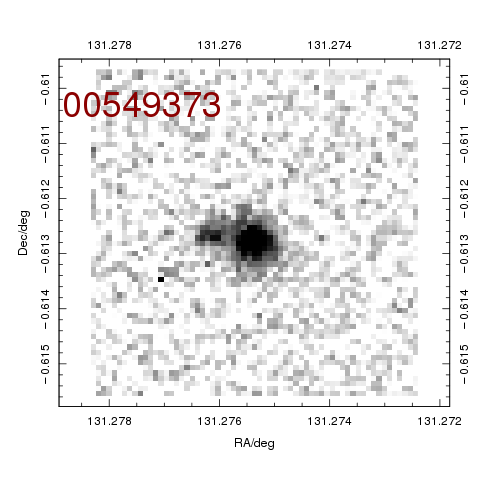}
\includegraphics[width=0.49\textwidth]{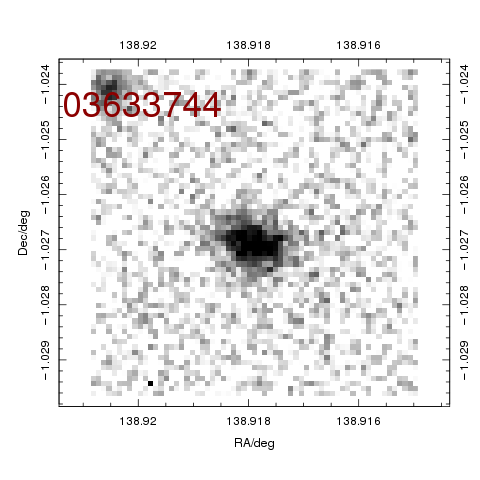}\\
\includegraphics[width=0.49\textwidth]{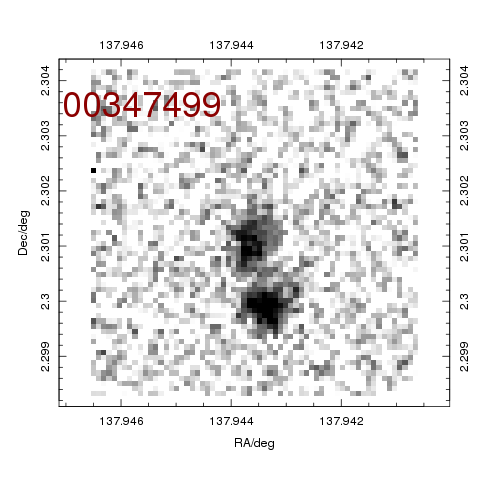}
\includegraphics[width=0.49\textwidth]{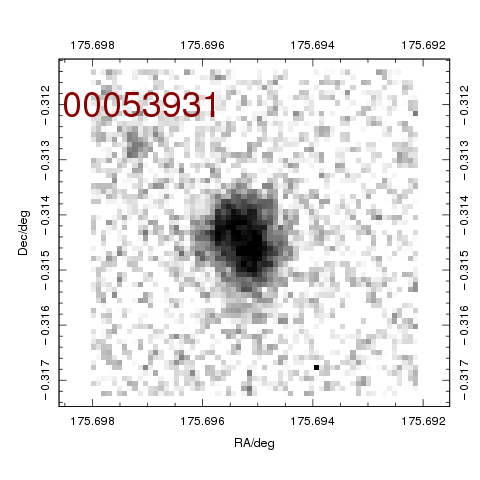}\\
\caption{-- {\em continued}}
\end{center}
\end{figure*}

%

\section{Concluding Remarks}
\label{s:disc}

From the $\sim$230\,000 objects with spectroscopy in the GAMA survey, we identified \nsamp\ blended objects ($\sim$0.12\%). 
In contrast, out of the 849\,920 spectra in SDSS/DR4 \citep{SDSS-DR4}, \cite{slacs5} identified a total of 89 lenses and \cite{Holwerda07c} identified 101 candidate occulting galaxy pairs, i.e., 0.02\% of all the SDSS spectra were blends. 
To make an honest comparison, we can only count the early-type, passive spectra with emission line at different redshift, the target of \cite{Bolton04} and the subsequent SLACS survey. In GAMA, we identify 104+71=179 of these (0.08\%), a factor four higher detection rate. 


There are obvious differences between the SDSS and GAMA redshift surveys: a fainter limiting magnitude and a higher completeness (by design) for the GAMA survey.
The fainter depth by two magnitudes means that there are about ten times as many galaxies with a similar magnitude
(e.g., $r\sim19.5$ for GAMA, $r\sim17.5$ for SDSS main galaxy sample).
This will result in more blended spectra despite the fact the AAOmega apertures are $2''$ compared to $3''$ for SDSS. This latter difference diminishes the GAMA survey's sensitivity to overlapping pairs; a wider aperture includes more flux from the outskirts of the occulting galaxy. Naively, one would therefore expect a factor $\sim$4.5 improvement in sensitivity of GAMA with respect to SDSS for blended spectra, which bears out approximately (0.12\% of GAMA vs. 0.02\% of SDSS/DR4), better than the increase in blended spectra from SDSS DR4 to DR10.
Another difference between the surveys is the identification of the redshift blend. In the SLACS survey, 
potentially lensed star-forming galaxies are detected through the presence of background oxygen and hydrogen nebular emission lines in the SDSS-DR4 spectra of massive foreground galaxies. 
The GAMA identification, this paper, is through a complete-spectrum cross-correlation with different templates, which uses the full spectral range to identify redshift, and allows for two blended passive spectra or at least does not require very strong emission lines at different redshifts.

There are several possible uses for these blended spectral galaxy pairs. 

The occulting pairs in GAMA are added to the master occulting galaxy catalogue, predominantly based on the SDSS spectral identifications \citep[86 blended pairs in][]{Holwerda07c}, and the GalaxyZoo identifications \cite[1993,][]{Keel13}. 

The presented catalogue of occulting pairs constitutes one way to identify occulting pairs in GAMA. Another approach uses the rejects from the pairs and group catalogue: galaxy pairs that are close on the sky but separated enough to warrant separate fibre assignment and do not exhibit a blended spectrum. By requiring that both pair members are well-separated in redshift, we obtain bona-fide occulting pairs.

A complete catalogue of galaxy groups is one of the primary goals of the GAMA survey \citep[][]{Robotham11,Robotham14}. A second sample of overlapping pairs will be identified from this catalogue, once it is complete ($\sim300$ expected). Therefore, between the high-fidelity automated identification of shared-fibre pairs and simultaneously, a complete census of close, serendipitous overlaps with separate redshifts, the GAMA identifications of overlapping galaxies will be the most complete to date. 

In the case of the lenses, the presented lensing galaxy candidates represent a near doubling of the known objects from SLACS (89 lenses) and a useful addition to the BOSS identified ones \citep{Bolton12a}.
The increased depth and completeness of GAMA means more distant and lower mass lenses are included. 
It should be illustrative to study these blended objects with the ongoing IFU surveys (e.g., SAMI or MANGA) and perhaps spatially separate the blended spectral signal or at least study the variation with fibre of the blended signal. For a full lensing analysis, the imaging will have to be higher spatial resolution than those available from either SDSS, KiDS or any of the other imaging surveys available for GAMA. Either dedicated VLT/AO observations or HST imaging would fit the bill. The benefits of the GAMA selection are lower-mass lenses and lensed images closer to the lensing galaxy.

\section*{Acknowledgements}

The authors thank the referee for his or her comments and suggestions.
The lead author thanks the European Space Agency for the support of the 
Research Fellowship program and the whole GAMA team for a magnificent observational effort.
GAMA is a joint European-Australasian project based around a spectroscopic campaign using the Anglo-Australian Telescope. The GAMA
input catalogue is based on data taken from the Sloan Digital Sky Survey and the UKIRT Infrared Deep Sky Survey. 
Complementary imaging of the GAMA regions is being obtained by a number of independent survey programs
including GALEX MIS, VST KiDS, VISTA VIKING,WISE, Herschel-ATLAS, GMRT and ASKAP providing
UV to radio coverage. GAMA is funded by the STFC (UK), the ARC (Australia), the AAO, and the participating
institutions. The GAMA website is \url{www.gamasurvey.org/}.
MJIB acknowledges financial support from the Australian Research Council (FT100100280)
This research has made use of the NASA/IPAC Extragalactic Database (NED) which is operated by the Jet Propulsion Laboratory, California Institute of Technology, under contract with the National Aeronautics and Space Administration. 
This research has made use of NASA's Astrophysics Data System.



\clearpage
\newpage

\begin{table*}
\caption{The complete catalogue of blended spectra in the GAMA survey (online in MNRAS). T1 and T2 refer to the template numbers for the first and second peaks. }
\begin{center}
\begin{tabular}{r r r r r r r r r c c c}
Field & GAMA-id & RA & DEC & $z$ & T1 & $r_x$ & $z_2$ & T2 & $r_{x,2}$ & Spec.\ Type & Vis.\ Type \\
\hline
\hline
 G09&  196060& 129.01621&  -0.69336& 0.293&40&  8.7& 0.051&46&  5.6& ELG+PG & SE    \\
 G09&  197073& 133.78179&  -0.74790& 0.270&40& 10.8& 0.268&44&  6.4& ELG+PG & EE    \\
 G09&  198082& 138.28150&  -0.66673& 0.163&40& 11.1& 0.321&47& 10.2& PG+ELG & ES    \\
 G09&  202448& 129.69546&  -0.38179& 0.418&40&  9.0& 0.738&45&  5.0& PG+ELG & SE    \\
 G09&  204140& 136.63883&  -0.35203& 0.282&40&  9.4& 0.449&47&  8.1& PG+ELG & Ef    \\
 G09&  209222& 132.36771&   0.16360& 0.128&40& 10.3& 0.603&47&  6.7& PG+ELG & E     \\
 G09&  209263& 132.50596&   0.04250& 0.310&42&  6.5& 0.270&46&  5.3& ELG+PG & Ef    \\
 G09&  209295& 132.61013&   0.11972& 0.313&40& 11.2& 0.608&47&  7.8& PG+ELG & Ef    \\
 G09&  209584& 134.02979&   0.15244& 0.167&46& 12.0& 0.158&40&  7.9& PG+ELG & SE    \\
 G09&  218757& 140.75621&   0.84809& 0.238&47&  9.9& 0.250&44&  9.1& ELG+ELG& Ef    \\
 G09&  279369& 136.79904&   0.98044& 0.284&40&  8.5& 0.238&46&  7.2& ELG+PG & ES/Ef \\
 G09&  279956& 140.14187&   0.97341& 0.586&47&  6.0& 0.336&40&  5.9& PG+ELG & ES    \\
 G09&  300500& 129.87425&   1.12844& 0.153&44&  9.5& 0.158&40&  7.4& ELG+PG & SS/Q  \\
 G09&  300979& 131.46746&   1.19884& 0.146&40&  6.8& 0.716&40&  5.8& PG+PG  & SE    \\
 G09&  301818& 135.32929&   1.22984& 0.487&47& 10.6& 0.247&40& 10.0& PG+ELG & SE/M  \\
 G09&  302123& 136.46358&   1.31167& 0.470&40&  6.7& 0.646&40&  6.3& PG+PG  & E     \\
 G09&  302719& 138.94058&   1.33144& 0.593&47&  6.5& 0.404&42&  6.4& PG+ELG & SS    \\
 G09&  323200& 130.73717&   1.55957& 0.416&47&  9.7& 0.350&40&  6.7& PG+ELG & EE    \\
 G09&  323247& 131.02333&   1.64814& 0.164&40& 10.6& 0.229&46&  8.5& PG+ELG & SE    \\
 G09&  324470& 136.07796&   1.68015& 0.148&45& 10.3& 0.321&42&  8.9& ELG+PG & SE/Phi\\
 G09&  325026& 138.25700&   1.72137& 0.329&40&  7.1& 0.336&45&  6.3& PG+ELG & ES/Ef \\
 G09&  325108& 138.62475&   1.87141& 0.202&46&  9.1& 0.165&40&  7.0& PG+ELG & Ef    \\
 G09&  345974& 131.75004&   2.04437& 0.207&46& 11.2& 0.231&45&  8.5& ELG+ELG& Ef    \\
 G09&  345984& 131.78108&   2.12672& 0.207&40&  8.1& 0.138&46&  7.6& ELG+PG & ES    \\
 G09&  347499& 137.94367&   2.30107& 0.316&45&  8.9& 0.381&47&  5.9& ELG+ELG& SS    \\
 G09&  348065& 140.09571&   2.33430& 0.281&45&  9.9& 0.086&45&  9.4& ELG+ELG& SS/M  \\
 G09&  348299& 140.94438&   2.24256& 0.409&42& 10.4& 0.279&45&  6.3& ELG+PG & SE    \\
 G09&  371208& 130.50925&   1.02071& 0.077&40&  7.7& 0.699&46&  6.0& PG+ELG & SS    \\
 G09&  371995& 133.78987&   0.98418& 0.108&47&  8.6& 0.386&42&  7.9& ELG+PG & M     \\
 G09&  372810& 137.79708&   1.07694& 0.301&40&  8.9& 0.220&45&  6.4& ELG+PG & ES    \\
 G09&  375506& 129.51933&   1.39028& 0.240&46& 13.5& 0.419&46&  7.5& ELG+ELG& SE/F  \\
 G09&  376767& 134.88888&   1.50709& 0.727&47&  7.4& 0.176&46&  6.7& ELG+ELG& SS    \\
 G09&  376791& 135.11008&   1.44247& 0.198&40& 10.4& 0.149&46&  9.1& ELG+PG & ES/M  \\
 G09&  377212& 136.80625&   1.64392& 0.283&40& 12.3& 0.308&44&  5.6& PG+ELG & EE/M  \\
 G09&  377486& 137.90642&   1.62916& 0.169&40& 10.7& 0.334&46&  5.0& PG+ELG & E     \\
 G09&  380839& 130.27629&   1.72911& 0.361&42&  8.4& 0.387&47&  7.3& PG+ELG & EE    \\
 G09&  382083& 135.24921&   1.89325& 0.098&47&  8.5& 0.217&40&  7.1& ELG+PG & ES/M  \\
 G09&  382857& 138.76350&   2.04159& 0.341&40&  6.1& 0.358&46&  4.9& PG+ELG & ES    \\
 G09&  383053& 139.68454&   1.98081& 0.189&45&  9.0& 0.671&47&  7.6& ELG+ELG& S     \\
 G09&  383284& 140.65750&   2.13736& 0.194&40&  9.7& 0.342&47&  5.8& PG+ELG & ES    \\
 G09&  386427& 132.24292&   2.24817& 0.209&42&  7.5& 0.393&40&  6.6& PG+PG  & EE    \\
 G09&  387615& 137.19829&   2.40849& 0.281&40& 10.5& 0.549&42&  5.1& PG+PG  & E/Ef  \\
 G09&  388201& 138.99892&   2.51429& 0.167&40& 10.0& 0.299&44&  8.2& PG+ELG & ES    \\
 G09&  417645& 132.51888&   2.29438& 0.108&40& 12.4& 0.457&45&  5.8& PG+ELG & S     \\
 G09&  417922& 134.09700&   2.49636& 0.198&40&  9.1& 0.615&42&  6.5& PG+PG  & Ef    \\
 G09&  419678& 140.98671&   2.86084& 0.569&40&  8.2& 0.662&47&  7.4& PG+ELG & E     \\
 G09&  422882& 132.81062&   2.83784& 0.308&45&  9.4& 0.212&40&  7.8& PG+ELG & SS    \\
 G09&  425637& 129.15721&   2.86991& 0.511&42&  8.0& 0.282&40&  6.1& PG+PG  & Ef    \\
 G09&  528021& 140.98654&  -0.87061& 0.157&40&  9.1& 0.136&44&  6.7& ELG+PG & ES/Ef \\
 G09&  573657& 129.41362&  -0.01039& 0.274&40&  7.2& 0.657&46&  6.8& PG+ELG & E     \\
 G09&  573744& 129.95100&  -0.03694& 0.144&45& 11.5& 0.130&45&  9.2& ELG+ELG& SE/B  \\
 G09&  574390& 135.66788&  -0.13742& 0.223&40& 11.6& 0.229&46&  5.5& PG+ELG & X     \\
 G09&  574857& 137.52142&  -0.12759& 0.105&45& 12.1& 0.300&44&  9.4& ELG+ELG& M/SS  \\
 G09&  575653& 140.77658&  -0.11463& 0.320&40& 11.0& 0.481&47&  6.1& PG+ELG & Ef    \\
 G09&  599598& 132.08975&   0.21503& 0.118&45&  6.4& 0.195&40&  5.8& ELG+PG & SE    \\
 G09&  599770& 132.67583&   0.34792& 0.191&40&  7.7& 0.333&40&  7.6& PG+PG  & SE    \\
 G09&  599797& 132.62096&   0.25907& 0.200&40&  9.3& 0.054&45&  7.9& ELG+PG & Ef    \\
 G09&  599995& 133.35229&   0.32283& 0.289&43&  9.1& 0.292&43&  5.6& ELG+ELG& E     \\
 G09&  600968& 138.08767&   0.30248& 0.053&46& 10.9& 0.316&42&  8.2& ELG+PG & SE/Phi\\
 G09&  621991& 129.85479&   0.68659& 0.145&40&  7.2& 0.078&45&  5.8& ELG+PG & SE    \\
 G09&  622326& 132.66296&   0.63611& 0.378&42&  7.5& 0.234&40&  7.1& PG+PG  & EE/Ef \\
\hline 
\end{tabular}
\end{center}
\end{table*}%

\begin{table*}
\setcounter{table}{3}
\caption{-- {\em continued}.}
\begin{center}
\begin{tabular}{r r r r r r r r r c c c}
Field & GAMA-id & RA & DEC & $z$ & T1 & $r_x$ & $z_2$ & T2 & $r_{x,2}$ & Spec.\ Type & Vis.\ Type \\
\hline
\hline
 G09&  641911& 140.01479&  -1.21574& 0.096&46&  9.9& 0.220&45&  9.1& ELG+ELG& SE    \\
 G09&  663372& 140.42683&  -1.09201& 0.331&45&  6.6& 0.224&40&  6.4& PG+ELG & Ef    \\
 G09&  905359& 139.63437&  -0.60728& 0.585&40&  6.5& 0.757&47&  6.2& PG+ELG & E     \\
 G09& 3576902& 129.95892&  -1.63319& 0.268&45&  9.7& 0.187&40&  7.3& PG+ELG & M     \\
 G09& 3579777& 131.41271&  -1.66217& 0.258&40&  7.9& 0.869&42&  5.3& PG+PG  & Ef    \\
 G09& 3612009& 135.81950&  -1.34174& 0.313&46&  7.2& 0.452&40&  6.1& ELG+PG & M     \\
 G09& 3624490& 132.09767&  -0.85274& 0.366&40&  8.3& 0.246&46&  7.4& ELG+PG & M/Ef  \\
 G09& 3626956& 134.23367&  -0.80412& 0.736&47&  6.6& 0.267&42&  6.5& PG+ELG & Ef    \\
 G09& 3632403& 138.10708&  -0.95852& 0.315&40& 10.1& 0.444&47&  8.5& PG+ELG & Ef    \\
 G09& 3633744& 138.91800&  -1.02683& 0.343&45&  7.4& 0.440&46&  5.1& ELG+ELG& ES    \\
 G09& 3866435& 136.82975&  -1.99357& 0.325&47&  9.3& 0.166&44&  8.5& ELG+ELG& S     \\
 G09& 3884455& 135.10842&  -1.60189& 0.411&42&  9.2& 0.440&47&  7.4& PG+ELG & Ef    \\
 G09& 3886423& 136.26788&  -1.56631& 0.224&47& 13.5& 0.284&40&  7.2& ELG+PG & ES    \\
 G09& 3887215& 136.64988&  -1.59199& 0.328&46&  7.0& 0.331&47&  6.3& ELG+ELG& SS    \\
 G09& 3890035& 138.55021&  -1.68663& 0.259&40&  9.1& 0.868&42&  6.0& PG+PG  & Ef    \\
 G09& 3904099& 134.01279&  -0.98051& 0.217&46& 11.5& 0.242&47&  6.5& ELG+ELG& SE    \\
 G09& 3909630& 137.37821&  -1.17997& 0.227&40& 10.2& 0.328&44&  8.2& PG+ELG & SS/M  \\
 G09& 3913313& 139.74904&  -1.30614& 0.223&46& 11.1& 0.226&40& 10.3& ELG+PG & E     \\
 G12&    7242& 176.29229&   0.73915& 0.529&40&  8.3& 0.642&47&  8.2& PG+ELG & E     \\
 G12&    7907& 179.68421&   0.82128& 0.230&46&  9.1& 0.244&40&  8.1& ELG+PG & SE    \\
 G12&    9156& 185.08025&   0.71944& 0.270&40& 10.1& 0.115&46&  8.9& ELG+PG & ES    \\
 G12&   22278& 176.96358&   1.20310& 0.094&40&  8.5& 0.712&47&  6.0& PG+ELG & S     \\
 G12&   22559& 178.05237&   1.05246& 0.283&45&  8.8& 0.157&40&  7.7& PG+ELG & SE    \\
 G12&   32248& 183.92221&  -1.20294& 0.085&47& 14.9& 0.328&45&  7.5& ELG+ELG& SE    \\
 G12&   39055& 174.87917&  -0.66397& 0.096&44&  5.2& 0.439&40&  5.0& ELG+PG & E     \\
 G12&   39870& 178.62517&  -0.72424& 0.312&40& 10.1& 0.227&40&  6.5& PG+PG  & SE    \\
 G12&   40642& 182.30704&  -0.78791& 0.278&47& 12.2& 0.184&45&  8.6& ELG+ELG& M     \\
 G12&   53931& 175.69525&  -0.31443& 0.121&47& 13.1& 0.147&45& 12.2& ELG+ELG& SS/F  \\
 G12&   54130& 176.85742&  -0.30877& 0.326&45& 11.6& 0.317&47&  8.6& ELG+ELG& SE/F  \\
 G12&   54523& 178.36975&  -0.30537& 0.260&47& 10.9& 0.077&46& 10.0& ELG+ELG& M     \\
 G12&   55091& 180.44258&  -0.34558& 0.239&40&  7.6& 0.169&45&  7.5& ELG+PG & ES    \\
 G12&   55175& 180.74654&  -0.38792& 0.357&47&  7.0& 0.409&47&  6.7& ELG+ELG& SE    \\
 G12&   71072& 182.90017&   0.17959& 0.255&45&  6.2& 0.122&40&  5.6& PG+ELG & SS    \\
 G12&   71566& 185.16637&   0.20776& 0.287&40& 11.5& 0.259&47&  6.5& ELG+PG & Phi   \\
 G12&   84090& 175.97379&   0.48608& 0.258&44&  8.2& 0.284&42&  7.2& ELG+PG & SS/M/F\\
 G12&   97836& 174.84862&   0.90228& 0.241&40& 10.9& 0.110&47&  8.7& ELG+PG & E     \\
 G12&   98399& 177.78321&   0.91949& 0.400&40&  8.9& 0.569&47&  7.3& PG+ELG & SE    \\
 G12&   98624& 179.06704&   0.99197& 0.259&40& 11.7& 0.270&47&  5.5& PG+ELG & E     \\
 G12&   99508& 183.13442&   1.00244& 0.214&40&  9.5& 0.744&43&  5.0& PG+ELG & M     \\
 G12&  124295& 178.90833&  -2.56675& 0.280&47& 11.2& 0.179&44& 10.2& ELG+ELG& M     \\
 G12&  125217& 183.07617&  -2.50805& 0.224&46&  7.4& 0.265&40&  6.0& ELG+PG & E     \\
 G12&  130303& 176.78942&  -2.18512& 0.435&47&  7.7& 0.398&40&  7.5& PG+ELG & SE    \\
 G12&  130523& 177.73192&  -2.10683& 0.163&47& 10.9& 0.249&44&  9.6& ELG+ELG& SE    \\
 G12&  131725& 182.36783&  -2.24046& 0.630&46&  9.7& 0.179&40&  8.2& PG+ELG & E     \\
 G12&  136782& 176.11142&  -1.77261& 0.394&40&  7.0& 0.246&46&  6.8& ELG+PG & SE    \\
 G12&  136800& 176.05287&  -1.77459& 0.264&40& 11.7& 0.589&47&  7.8& PG+ELG & M     \\
 G12&  136907& 176.30279&  -1.74244& 0.028&47& 11.3& 0.256&45&  7.6& ELG+ELG& SE    \\
 G12&  137783& 179.51112&  -1.69882& 0.097&47& 10.9& 0.268&45&  8.8& ELG+ELG& M/SS  \\
 G12&  138015& 180.50629&  -1.65407& 0.267&40& 10.9& 0.250&47&  5.2& ELG+PG & E     \\
 G12&  138368& 181.94588&  -1.82568& 0.304&45&  8.3& 0.274&47&  6.3& ELG+ELG& E     \\
 G12&  138811& 184.17454&  -1.81681& 0.450&47&  8.6& 0.074&45&  5.9& ELG+ELG& S     \\
 G12&  164640& 178.14021&  -2.71487& 0.078&40& 11.6& 0.153&45&  6.7& PG+ELG & SS    \\
 G12&  164995& 179.40017&  -2.87571& 0.267&45&  9.5& 0.270&46&  8.0& ELG+ELG& Q     \\
 G12&  172153& 181.44675&  -2.44182& 0.514&43&  6.2& 0.567&42&  5.5& ELG+PG & SE    \\
 G12&  177278& 176.08696&  -1.86907& 0.316&43&  5.8& 0.385&45&  5.1& ELG+ELG& M     \\
 G12&  184530& 176.70954&  -1.43849& 0.391&40& 10.3& 0.156&45&  9.7& ELG+PG & SE    \\
 G12&  185604& 180.96967&  -1.46278& 0.265&40& 10.6& 0.273&47&  9.5& PG+ELG & SE/F  \\
 G12&  185812& 181.78058&  -1.54840& 0.313&47& 10.1& 0.181&46&  9.4& ELG+ELG& ES    \\
 G12&  185998& 182.60063&  -1.61024& 0.414&40&  9.0& 0.288&45&  7.4& ELG+PG & ES    \\
 G12&  186085& 182.89554&  -1.64913& 0.241&45&  7.6& 0.380&40&  7.0& ELG+PG & E     \\
 G12&  186516& 184.85638&  -1.53982& 0.324&40&  9.2& 0.265&45&  5.8& ELG+PG & SE/F  \\
 G12&  186737& 185.59350&  -1.49532& 0.301&43&  8.1& 0.297&45&  5.9& ELG+ELG& M     \\
\hline 
\end{tabular}
\end{center}
\end{table*}%

\begin{table*}
\setcounter{table}{3}
\caption{ -- {\em continued}.}
\begin{center}
\begin{tabular}{r r r r r r r r r c c c}
Field & GAMA-id & RA & DEC & $z$ & T1 & $r_x$ & $z_2$ & T2 & $r_{x,2}$ & Spec.\ Type & Vis.\ Type \\
\hline
\hline
 G12&  220204& 180.67683&   1.56371& 0.353&42&  8.0& 0.228&47&  6.5& ELG+PG & ES    \\
 G12&  220682& 182.74350&   1.59787& 0.289&40&  9.8& 0.361&47&  9.8& PG+ELG & Ef    \\
 G12&  220854& 183.53083&   1.55306& 0.266&40&  9.6& 0.327&46&  9.2& PG+ELG & SE    \\
 G12&  230803& 181.08337&   1.93953& 0.166&44& 10.2& 0.243&45&  9.2& ELG+ELG& ES    \\
 G12&  231043& 182.16688&   1.99840& 0.160&40& 10.1& 0.689&47&  8.9& PG+ELG & Ef    \\
 G12&  231307& 183.15171&   1.97018& 0.190&40& 12.2& 0.218&47&  6.7& PG+ELG & Ef    \\
 G12&  231746& 185.33296&   1.92256& 0.423&40&  9.5& 0.460&47&  6.9& PG+ELG & ES/Ef \\
 G12&  231768& 185.44796&   1.94621& 0.226&44& 11.3& 0.252&47&  8.7& ELG+ELG& ES    \\
 G12&  231785& 185.53046&   1.88483& 0.112&44&  9.1& 0.346&47&  6.4& ELG+ELG& M     \\
 G12&  272227& 178.47396&   1.32070& 0.078&47& 10.2& 0.230&44& 10.1& ELG+ELG& SE    \\
 G12&  272828& 180.97808&   1.42330& 0.402&40&  9.9& 0.238&47&  6.9& ELG+PG & Phi   \\
 G12&  273903& 185.65375&   1.33724& 0.217&40& 10.1& 0.415&46&  6.6& PG+ELG & E     \\
 G12&  289050& 180.98801&   1.78262& 0.262&43& 12.5& 0.259&45&  6.6& ELG+ELG& M     \\
 G12&  289265& 181.63046&   1.82378& 0.295&40&  9.1& 0.292&40&  7.3& PG+PG  & ES    \\
 G12&  289278& 181.64301&   1.74123& 0.294&42& 10.9& 0.701&45&  5.0& PG+ELG & ES    \\
 G12&  396570& 174.04408&   1.44452& 0.454&47&  8.1& 0.077&44&  6.7& ELG+ELG& S/M   \\
 G12&  397097& 176.71996&   1.48606& 0.156&40&  8.5& 0.152&44&  6.5& ELG+PG & SE/F  \\
 G12&  402278& 174.23017&   1.89732& 0.138&40& 10.2& 0.204&45&  5.7& PG+ELG & SS    \\
 G12&  537363& 184.95279&  -0.87211& 0.040&46&  7.5& 0.206&40&  7.2& ELG+PG & SE/B  \\
 G12&  559147& 175.88608&  -0.43995& 0.351&40&  9.9& 0.803&42&  5.4& PG+PG  & EE    \\
 G12&  559653& 177.75950&  -0.47629& 0.124&44& 10.1& 0.259&44&  9.4& ELG+ELG& SS/F  \\
 G12&  560010& 178.85871&  -0.57989& 0.225&47& 12.0& 0.054&47&  9.7& ELG+ELG& SS/M  \\
 G12&  560475& 180.41129&  -0.48254& 0.165&40&  9.9& 0.543&47&  6.5& PG+ELG & EE    \\
 G12&  560853& 181.98808&  -0.52788& 0.224&40&  7.1& 0.682&47&  5.4& PG+ELG & Ef    \\
 G12&  583469& 175.01863&  -0.14991& 0.314&40& 11.2& 0.747&47&  7.4& PG+ELG & Ef    \\
 G12&  583813& 176.83237&  -0.16391& 0.137&45&  7.3& 0.109&47&  5.9& ELG+ELG& Ef    \\
 G12&  585446& 182.74083&  -0.13542& 0.345&42&  7.5& 0.331&40&  6.3& PG+PG  & SE    \\
 G12&  585644& 183.83021&  -0.19625& 0.404&40&  9.4& 0.256&46&  7.8& ELG+PG & EE/Ef \\
 G12&  586392& 184.62887&  -0.18264& 0.456&45&  7.8& 0.307&40&  6.5& PG+ELG & Ef    \\
 G12&  586648& 185.72429&  -0.20906& 0.697&47& 10.3& 0.287&40&  9.7& PG+ELG & Ef    \\
 G12&  610029& 178.25696&   0.30019& 0.365&40&  9.0& 0.391&47&  5.7& PG+ELG & SE    \\
 G12&  610529& 180.54204&   0.39134& 0.256&45& 11.3& 0.701&47&  7.3& ELG+ELG& Ef    \\
 G12&  690020& 179.39604&  -1.13984& 0.263&40&  8.8& 0.348&47&  6.4& PG+ELG & ES/Q  \\
 G12&  746445& 179.08296&  -0.20491& 0.107&40&  9.7& 0.645&42&  5.5& PG+PG  & EE    \\
 G12&  787196& 177.01971&  -1.87064& 0.323&47&  9.3& 0.116&44&  7.5& ELG+ELG& M     \\
 G12&  814991& 181.52958&   1.93496& 0.223&40&  8.3& 0.297&44&  5.3& PG+ELG & Ef    \\
 G12&  947988& 177.90113&  -1.11874& 0.295&44&  7.4& 0.136&45&  6.3& ELG+ELG& SS    \\
 G15&   14983& 213.63450&   0.74998& 0.273&40& 10.4& 0.528&47&  9.5& PG+ELG & EE    \\
 G15&   15221& 214.61879&   0.62943& 0.232&40& 10.9& 0.279&47&  7.7& PG+ELG & SE    \\
 G15&   15554& 215.97479&   0.67592& 0.142&40&  7.7& 0.227&46&  6.4& PG+ELG & SE    \\
 G15&   17244& 222.50238&   0.81555& 0.447&40&  7.2& 0.496&42&  4.7& PG+PG  & SE    \\
 G15&   48883& 218.90542&  -0.74389& 0.343&46&  9.1& 0.245&40&  8.6& PG+ELG & SE    \\
 G15&   62459& 213.06783&  -0.25768& 0.187&44&  7.7& 0.424&40&  6.5& ELG+PG & E     \\
 G15&   63082& 214.46454&  -0.30878& 0.251&47& 12.3& 0.124&45&  8.8& ELG+ELG& SS/B  \\
 G15&   64181& 218.48892&  -0.31288& 0.403&40&  7.5& 0.214&46&  7.4& ELG+PG & ES    \\
 G15&   65016& 221.17254&  -0.29001& 0.239&46& 12.9& 0.286&46& 11.0& ELG+ELG& S     \\
 G15&   65298& 222.49875&  -0.40713& 0.253&40&  8.7& 0.649&47&  8.3& PG+ELG & E     \\
 G15&   77892& 215.20808&   0.17706& 0.316&40&  9.0& 0.620&45&  5.7& PG+ELG & E     \\
 G15&   78841& 218.90942&   0.12274& 0.218&40& 12.1& 0.215&47&  5.1& ELG+PG & ES    \\
 G15&   92523& 216.62879&   0.62516& 0.125&46&  7.9& 0.100&47&  7.5& ELG+ELG& S     \\
 G15&  106821& 217.32700&   0.86711& 0.160&40&  9.8& 0.452&47&  6.2& PG+ELG & E     \\
 G15&  238715& 215.06083&   1.60423& 0.190&40& 11.0& 0.243&45&  9.3& PG+ELG & M     \\
 G15&  239143& 216.51021&   1.60820& 0.084&45&  8.6& 0.294&44&  5.9& ELG+ELG& S     \\
 G15&  239367& 217.45604&   1.52598& 0.284&40& 10.2& 0.379&45&  8.5& PG+ELG & ES    \\
 G15&  239721& 218.71629&   1.60518& 0.199&47&  9.6& 0.056&47&  8.9& ELG+ELG& SE/B  \\
 G15&  240564& 223.16475&   1.38699& 0.213&45&  9.3& 0.275&47&  6.2& ELG+ELG& SE/Ef \\
 G15&  249903& 213.33275&   2.02850& 0.176&47&  9.7& 0.179&43&  7.1& ELG+ELG& S     \\
 G15&  250209& 214.17046&   2.07386& 0.553&47&  7.1& 0.386&40&  6.6& PG+ELG & Ef    \\
 G15&  250487& 214.82401&   2.12522& 0.304&44&  7.1& 0.269&40&  5.9& PG+ELG & EE    \\
 G15&  250796& 216.02125&   2.13975& 0.199&40& 10.2& 0.132&45&  6.4& ELG+PG & Ef    \\
 G15&  251342& 218.22696&   1.97782& 0.439&40&  7.1& 0.466&43&  5.8& PG+ELG & EE    \\
 G15&  251526& 219.06642&   1.97840& 0.063&46&  8.5& 0.283&45&  7.5& ELG+ELG& S/M   \\
\hline 
\end{tabular}
\end{center}
\end{table*}%

\begin{table*}
\setcounter{table}{3}
\caption{ -- {\em continued}.}
\begin{center}
\begin{tabular}{r r r r r r r r r c c c}
Field & GAMA-id & RA & DEC & $z$ & T1 & $r_x$ & $z_2$ & T2 & $r_{x,2}$ & Spec.\ Type & Vis.\ Type \\
\hline
\hline
 G15&  262626& 220.31992&   2.41527& 0.287&40&  6.9& 0.028&44&  6.0& ELG+PG & SE/B  \\
 G15&  262987& 222.23792&   2.38034& 0.562&42&  7.6& 0.207&40&  7.1& PG+PG  & ES/Ef \\
 G15&  266240& 216.94617&   2.79181& 0.297&45&  6.1& 0.378&40&  5.3& ELG+PG & ES    \\
 G15&  278240& 215.80875&   1.07883& 0.393&44&  7.9& 0.282&47&  5.4& ELG+ELG& EE    \\
 G15&  296898& 213.92688&   1.53891& 0.160&40& 11.0& 0.045&47&  8.5& ELG+PG & M/Ef  \\
 G15&  297627& 216.43496&   1.38861& 0.453&47&  7.8& 0.197&47&  7.2& ELG+ELG& SS/M  \\
 G15&  297645& 216.57825&   1.38436& 0.316&47& 12.8& 0.004&44&  6.3& ELG+ELG& SE/M  \\
 G15&  298302& 219.30533&   1.41623& 0.403&40&  6.5& 0.150&40&  6.0& PG+PG  & EE    \\
 G15&  298316& 219.34801&   1.28769& 0.333&40&  7.4& 0.329&40&  6.8& PG+PG  & ES    \\
 G15&  298899& 222.14058&   1.16504& 0.319&40&  7.0& 0.374&47&  6.3& PG+ELG & E     \\
 G15&  319348& 214.14233&   1.86348& 0.266&40&  8.6& 0.286&46&  7.2& PG+ELG & SE    \\
 G15&  319416& 214.38538&   1.91140& 0.051&40&  8.3& 0.245&43&  6.5& PG+ELG & S     \\
 G15&  320078& 216.84646&   1.88291& 0.348&42&  7.5& 0.292&42&  7.5& PG+PG  & EE/M  \\
 G15&  320384& 218.13475&   1.68924& 0.109&40&  8.4& 0.192&46&  7.8& PG+ELG & SS    \\
 G15&  320557& 218.63421&   1.78566& 0.274&40& 11.1& 0.279&46&  7.2& PG+ELG & EE    \\
 G15&  320705& 219.18917&   1.74384& 0.176&44&  8.1& 0.546&47&  6.6& ELG+ELG& S     \\
 G15&  321208& 221.87658&   1.67057& 0.358&40&  7.8& 0.286&46&  5.2& ELG+PG & EE    \\
 G15&  342308& 215.06900&   2.22422& 0.391&45&  9.6& 0.328&40&  5.9& PG+ELG & SE    \\
 G15&  343868& 222.41704&   2.17379& 0.121&47& 14.4& 0.124&47&  7.9& ELG+ELG& SS    \\
 G15&  362394& 214.88958&   2.70961& 0.394&42&  7.9& 0.315&40&  6.4& PG+PG  & E/Ef  \\
 G15&  367197& 220.31267&   2.92665& 0.212&40&  6.8& 0.098&44&  5.9& ELG+PG & Ef    \\
 G15&  460386& 211.94129&  -1.80552& 0.113&46&  8.5& 0.516&43&  8.5& ELG+ELG& SE    \\
 G15&  460463& 212.48058&  -1.61746& 0.115&40& 10.8& 0.411&47&  6.8& PG+ELG & SE/Ef \\
 G15&  460713& 213.41046&  -1.64736& 0.429&47&  9.4& 0.308&40&  6.4& PG+ELG & ES    \\
 G15&  463638& 213.94213&  -1.18728& 0.050&47& 10.6& 0.149&47&  8.1& ELG+ELG& SE    \\
 G15&  485756& 217.11896&  -1.79492& 0.269&40& 11.5& 0.404&47&  6.2& PG+ELG & EE    \\
 G15&  493692& 221.98950&  -1.36708& 0.261&47& 13.0& 0.147&45&  8.5& ELG+ELG& SS    \\
 G15&  508116& 215.95117&  -1.68131& 0.202&40&  7.8& 0.305&44&  6.9& PG+ELG & ES/Ef \\
 G15&  508211& 216.27379&  -1.59829& 0.289&40&  8.4& 0.278&47&  6.7& ELG+PG & SE    \\
 G15&  512280& 218.33704&  -1.05040& 0.179&45&  9.2& 0.030&47&  8.8& ELG+ELG& SS/B  \\
 G15&  513306& 223.07412&  -1.01726& 0.211&46& 11.3& 0.320&42&  8.2& ELG+PG & ES/Ef \\
 G15&  543716& 212.59971&  -0.92718& 0.357&47&  8.3& 0.204&44&  6.6& ELG+ELG& SS    \\
 G15&  544875& 217.28529&  -0.94940& 0.394&42&  8.1& 0.494&40&  5.3& PG+PG  & EE    \\
 G15&  544985& 217.81129&  -0.90871& 0.402&40&  8.4& 0.400&45&  5.4& ELG+PG & Ef    \\
 G15&  545139& 218.38854&  -1.00978& 0.241&46&  6.5& 0.323&40&  6.0& ELG+PG & SE    \\
 G15&  545712& 220.81617&  -0.88778& 0.450&42&  9.4& 0.216&46&  7.2& ELG+PG & ES/M  \\
 G15&  570323& 223.22658&  -0.58658& 0.313&45& 13.3& 0.280&47&  5.5& ELG+ELG& ES/Ef \\
 G15&  592207& 212.01942&  -0.07878& 0.256&40&  8.3& 0.178&40&  7.3& PG+PG  & Ef    \\
 G15&  617535& 211.95117&   0.23064& 0.510&47&  8.0& 0.262&40&  6.0& PG+ELG & Ef    \\
 G15&  617856& 213.49346&   0.35214& 0.231&40& 11.7& 0.262&47&  6.7& PG+ELG & SE    \\
 G15&  618687& 216.50154&   0.23066& 0.150&47& 13.2& 0.084&47&  9.5& ELG+ELG& SS    \\
 G15&  884103& 217.42279&  -1.68879& 0.793&47&  8.5& 0.018&47&  7.4& ELG+ELG& SS    \\
 G02& 1126606&  33.89513&  -6.55139& 0.217&40&  9.8& 0.749&43&  5.1& PG+ELG & SE    \\
 G02& 1217811&  33.60950&  -4.23305& 0.181&40&  7.1& 0.154&44&  6.6& ELG+PG & Phi   \\
 G02& 1270758&  32.65929& -10.07633& 0.144&40&  9.7& 0.698&43&  5.6& PG+ELG & S     \\
 G02& 1274050&  32.82713&  -9.86427& 0.161&44&  6.9& 0.507&47&  6.5& ELG+ELG& SS/Phi\\
 G02& 1298084&  32.37917&  -6.37684& 0.238&47&  6.1& 0.273&40&  6.0& ELG+PG & EE    \\
 G02& 1312058&  32.29279&  -5.43011& 0.417&40&  8.2& 0.619&47&  7.1& PG+ELG & Ef    \\
 G02& 1320592&  32.94087&  -4.82006& 0.070&47& 11.7& 0.162&46& 10.0& ELG+ELG& SE    \\
 G02& 1440776&  31.98304&  -4.41341& 0.209&45&  6.3& 0.640&40&  5.9& ELG+PG & EE    \\
 G02& 1537351&  30.80412&  -6.04563& 0.424&40&  6.6& 0.622&42&  6.5& PG+PG  & EE    \\
 G02& 1568229&  30.69258&  -4.13571& 0.291&40&  5.9& 0.413&40&  5.8& PG+PG  & ES/Ef \\
 G02& 1569977&  30.69342&  -4.01222& 0.331&40& 10.3& 0.436&47&  5.3& PG+ELG & Ef    \\
 G02& 1614527&  35.88917&  -8.47218& 0.283&40& 11.6& 0.195&45&  5.1& ELG+PG & Ef/S  \\
 G02& 1675035&  35.41175&  -4.48197& 0.258&40&  7.7& 0.080&46&  5.8& ELG+PG & Ef/ES \\
 G02& 1684064&  35.62492&  -3.87688& 0.614&47&  8.4& 0.292&40&  6.6& PG+ELG & SE/M  \\
 G02& 1726233&  36.65454&  -8.52167& 0.184&47& 10.3& 0.155&47&  5.4& ELG+ELG& S     \\
 G02& 1740396&  36.36688&  -9.35231& 0.196&47& 12.3& 0.445&40&  6.3& ELG+PG & SE    \\
 G02& 1760310&  35.95496&  -6.11922& 0.236&45& 11.2& 0.144&45&  8.3& ELG+ELG& SS    \\
 G02& 1763319&  36.37075&  -5.91766& 0.294&40&  7.2& 0.319&44&  6.6& PG+ELG & SS    \\
 G02& 1765570&  36.16213&  -5.76612& 0.231&40&  7.4& 0.162&44&  6.3& ELG+PG & ES    \\
 G02& 1771132&  36.28483&  -5.39210& 0.297&46& 10.1& 0.231&44&  6.9& ELG+ELG& S/M   \\
 G02& 1779869&  36.29463&  -4.84248& 0.309&44&  6.2& 0.460&46&  5.8& ELG+ELG& EE    \\
 \hline 
\end{tabular}
\end{center}
\end{table*}%

\begin{table*}
\setcounter{table}{3}
\caption{ -- {\em continued}.}
\begin{center}
\begin{tabular}{r r r r r r r r r c c c}
Field & GAMA-id & RA & DEC & $z$ & T1 & $r_x$ & $z_2$ & T2 & $r_{x,2}$ & Spec.\ Type & Vis.\ Type \\
\hline
\hline
 G02& 1988308&  38.11142&  -5.96424& 0.300&47&  8.2& 0.303&47&  7.4& ELG+ELG& SS    \\
 G02& 2002618&  38.30588&  -5.07329& 0.348&42&  9.7& 0.256&45&  6.2& ELG+PG & SE    \\
 G02& 2005629&  37.92875&  -4.88681& 0.189&40&  6.9& 0.716&45&  5.9& PG+ELG & SE/S  \\
 G02& 2007752&  38.48246&  -4.76219& 0.355&47&  8.4& 0.034&44&  6.0& ELG+ELG& S     \\
 G02& 2248952&  30.82808&  -7.63584& 0.242&40&  5.4& 0.328&45&  5.2& PG+ELG & ES    \\
 G02& 2308869&  35.92083&  -4.02211& 0.237&47& 11.2& 0.264&47& 10.1& ELG+ELG& SS    \\
 G02& 2379807&  37.16025&  -4.03447& 0.333&40& 10.8& 0.329&46&  8.6& ELG+PG & Ef    \\
 G23& 5000687& 344.09237& -34.88054& 0.199&42&  8.7& 0.326&47&  7.5& PG+ELG & -     \\
 G23& 5006684& 343.38587& -33.71781& 0.219&40&  9.6& 0.208&47&  8.5& ELG+PG & -     \\
 G23& 5014213& 350.01962& -32.35935& 0.202&42& 10.3& 0.785&43&  6.0& PG+ELG & -     \\
 G23& 5022159& 347.44408& -31.04443& 0.213&40&  8.0& 0.235&45&  6.1& PG+ELG & -     \\
 G23& 5026858& 340.90525& -30.26776& 0.151&40&  8.4& 0.282&44&  5.5& PG+ELG & -     \\
 G23& 5027548& 341.34442& -30.14947& 0.282&46&  8.5& 0.311&47&  6.8& ELG+ELG& -     \\
 G23& 7014061& 343.38537& -33.71829& 0.208&46&  9.9& 0.224&40&  8.0& ELG+PG & -     \\
 G23& 7019633& 349.60625& -33.45856& 0.165&40&  8.7& 0.187&40&  8.3& PG+PG  & -     \\
 G23& 7021723& 345.68529& -33.36733& 0.221&40& 10.8& 0.319&47&  6.9& PG+ELG & -     \\
 G23& 7023086& 346.05408& -33.30514& 0.212&40&  7.8& 0.188&45&  7.0& ELG+PG & -     \\
 G23& 7028850& 342.77721& -33.04329& 0.321&40&  6.6& 0.328&47&  6.2& PG+ELG & -     \\
 G23& 7032116& 341.93799& -32.92324& 0.672&40&  7.0& 0.476&40&  6.5& PG+PG  & -     \\
 G23& 7035191& 346.47717& -32.81103& 0.634&43&  8.0& 0.188&40&  5.9& PG+ELG & -     \\
 G23& 7046392& 343.18829& -32.36551& 0.289&40&  9.3& 0.116&46&  6.0& ELG+PG & -     \\
 G23& 7048497& 342.86779& -32.27196& 0.239&45&  6.2& 0.389&45&  5.7& ELG+ELG& -     \\
 G23& 7056621& 342.57587& -31.90962& 0.388&40&  6.6& 0.425&47&  6.2& PG+ELG & -     \\
 G23& 7061088& 342.12237& -31.71619& 0.240&40&  7.7& 0.243&44&  6.6& PG+ELG & -     \\
 G23& 7069205& 342.01071& -31.37594& 0.261&46&  8.9& 0.432&40&  6.5& ELG+PG & -     \\
 G23& 7070361& 350.69167& -31.32483& 0.264&47&  7.3& 0.268&47&  5.5& ELG+ELG& -     \\
 G23& 7073988& 346.66158& -31.16843& 0.162&47& 12.4& 0.068&47&  8.9& ELG+ELG& -     \\
 G23& 7073990& 346.66237& -31.16836& 0.068&47& 12.2& 0.162&46&  7.8& ELG+ELG& -     \\
 G23& 7076874& 346.36892& -31.04973& 0.270&40&  8.4& 0.088&45&  7.5& ELG+PG & -     \\
 G23& 7081863& 340.21004& -30.82531& 0.287&45&  9.6& 0.217&40&  8.5& PG+ELG & -     \\
 G23& 7083868& 343.97800& -30.75893& 0.359&40&  6.7& 0.722&47&  6.6& PG+ELG & -     \\
 G23& 7093351& 344.23937& -30.36704& 0.449&40&  7.7& 0.325&47&  7.2& ELG+PG & -     \\
\hline 
\end{tabular}
\end{center}
\end{table*}%

\label{lastpage}

\end{document}